\documentclass[twocolumn,prd,showpacs,amsmath,amssymb]{revtex4}

\usepackage{graphicx}
\usepackage{dcolumn}
\usepackage{bm}
\usepackage{epsfig}
\usepackage{amssymb}
\usepackage{amsmath}

 \def\tskip{\setlength{\tskip}{5pt}}
\def\colwidth{\setlength{\colwidth}{3.5in}}

\newcommand{\aap}{Astron. Astrophys.}

\newcommand{\apjl}{Astrophys. J. Lett.}
\newcommand{\apjs}{Astrophys. J. Suppl. Ser.}

\newcommand{\arXiv}{ArXiv e-prints}
\newcommand{\azh}{Astron. Zh.}

\newcommand{\jcap}{JCAP}

\newcommand{\mnras}{Mon. Not. R. Astron. Soc.}

\newcommand{\plb}{Phys. Lett. B}

\newcommand{\lsim}{\mathrel{\hbox{\rlap{\lower.55ex\hbox{$\sim$}} \kern-.3em \raise.4ex \hbox{$<$}}}}
\newcommand{\gsim}{\mathrel{\hbox{\rlap{\lower.55ex\hbox{$\sim$}} \kern-.3em \raise.4ex \hbox{$>$}}}}
\newcommand{\beq}{\begin{equation}}
\newcommand{\eeq}{\end{equation}}
\newcommand{\beqa}{\begin{eqnarray}}
\newcommand{\eeqa}{\end{eqnarray}}
\newcommand{\drm}{\mathrm{d}}

\newcommand{\eh}{\epsilon_H}
\newcommand{\ev}{\epsilon_V}
\newcommand{\nh}{\eta_H}

\newcommand{\mpl}{m_\mathrm{Pl}}

\newcommand{\lcdm}{$\Lambda$CDM }

\newcommand{\et}{\tilde{\epsilon}}
\newcommand{\ximax}{\xi_\mathrm{max}}
\newcommand{\ximin}{\xi_\mathrm{min}}
\newcommand{\etmax}{\et_\mathrm{max}}
\newcommand{\etmin}{\et_\mathrm{min}}
\newcommand{\en}{\epsilon_N}
\newcommand{\ehmax}{\epsilon_{H,\mathrm{max}}}
\newcommand{\ehmin}{\epsilon_{H,\mathrm{min}}}
\newcommand{\smr}{S_\mathrm{m\gamma}}
\newcommand{\ssigr}{S_{\sigma\gamma}}

\newcommand{\zr}{\zeta_\gamma}
\newcommand{\zi}{\zeta^{(\mathrm{i})}}
\newcommand{\zf}{\zeta^{(\mathrm{f})}}

\newcommand{\trs}{{\cal T}_{\zeta S}}
\newcommand{\tss}{{\cal T}_{SS}}
\newcommand{\tilB}{\tilde{B}}
\newcommand{\Asqrd}{{\cal A}^2}
\newcommand{\Bsqrd}{{\cal B}^2}
\newcommand{\Cad}{\hat{C}_\ell^\mathrm{ad}}
\newcommand{\Ciso}{\hat{C}_\ell^\mathrm{iso}}
\newcommand{\Ccor}{\hat{C}_\ell^\mathrm{cor}}
\newcommand{\Atil}{\tilde{A}}

\begin{document}

\title{A Scale-Dependent Power Asymmetry from Isocurvature Perturbations}

\author{Adrienne L. Erickcek, Christopher M. Hirata, and Marc Kamionkowski}
\affiliation{Theoretical Astrophysics, California Institute of Technology, Mail Code
350-17, Pasadena, CA 91125}

\date{\today}

\begin{abstract}
If the hemispherical power asymmetry observed in the cosmic microwave background (CMB) on large angular scales is attributable to a superhorizon curvaton fluctuation, then the simplest model predicts that the primordial density fluctuations should be similarly asymmetric on all smaller scales.  The distribution of high-redshift quasars was recently used to constrain the power asymmetry on scales $k\simeq 1.5h$ Mpc$^{-1}$, and the upper bound on the amplitude of the asymmetry was found to be a factor of six smaller than the amplitude of the asymmetry in the CMB.  We show that it is not possible to generate an asymmetry with this scale dependence by changing the relative contributions of the inflaton and curvaton to the adiabatic power spectrum.  Instead, we consider curvaton scenarios in which the curvaton decays after dark matter freezes out, thus generating isocurvature perturbations.  If there is a superhorizon fluctuation in the curvaton field, then the rms amplitude of these perturbations will be asymmetric, and the asymmetry will be most apparent on large angular scales in the CMB.  We find that it is only possible to generate the observed asymmetry in the CMB while satisfying the quasar constraint if the curvaton's contribution to the total dark matter density is small, but nonzero.  The model also requires that the majority of the primordial power comes from fluctuations in the inflaton field.   Future observations and analyses of the CMB will test this model because the power asymmetry generated by this model has a specific spectrum, and the model requires that the current upper bounds on isocurvature power are nearly saturated.
\end{abstract}

\pacs{98.80.Cq, 98.70.Vc, 98.80.Bp}

\maketitle

%%%%%%%%%%%%%%%
\section{Introduction}
\label{sec:intro}
%%%%%%%%%%%%%%%%%%
The cosmic microwave background (CMB) \cite{BOOM,Miller99,Maxima2000,DASI02,CBI03,Archeops03,Acbar03,WMAP1cos,WMAP3,Acbar09,WMAP5only, WMAP5,Quad09,Bicep09} and the distribution of galaxies \cite{2dFpower, SDSS} tell us that the early Universe was homogeneous on superhorizon scales, spatially flat, and contained a nearly scale-invariant spectrum of adiabatic fluctuations.  These features of the early Universe provide compelling evidence for inflation \cite{Guth80, AS82, Linde81}.  Inflation also predicts that the observable Universe should be statistically isotropic; any anisotropy that may have existed prior to inflation would be stretched beyond the cosmological horizon during inflation.

There are indications, however, that the distribution of density perturbations is not isotropic \cite{TOH03, OTZH04, LM05, HBG04, Eriksen04, Eriksen07, HBGEL08, Eriksen09, DD05, Jaffe05, Jaffe06, Bielewicz05, CHSS06, Bernui06, GE08}.  In this article, we will focus our attention on one of these anomalies: the rms temperature fluctuation in the CMB on one side of the sky is larger than on the other side \cite{HBG04, Eriksen04, Eriksen07, HBGEL08, Eriksen09}.  This hemispherical power asymmetry can be parameterized as a dipolar modulation of the temperature anisotropy field \cite{Eriksen07, Eriksen09}; the temperature fluctuation in the $\hat{n}$ direction is
\beq
\frac{\delta T}{T}(\hat{n}) = s(\hat{n})\left[1+A(\hat{n}\cdot \hat{p})\right],
\label{defA}
\eeq
where $s(\hat{n})$ is an isotropic Gaussian random field.$^{1}$ \footnotetext[1]{This parameterization is based on a phenomenological model proposed in Ref. \cite{GHHC05}.}   The magnitude of the asymmetry is given by $A$ and its direction is specified by $\hat{p}$; the most recent analysis \cite{Eriksen09}, using the WMAP5 Internal Linear Combination (ILC) map \cite{ILC5yr}, found $A = 0.072 \pm 0.022$ for $\ell \lsim 64$ with $\hat{p}$ pointing at $(\ell, b) = (224^\circ, -22^\circ)\pm24^\circ$.  No explanation for the asymmetry involving foregrounds or systematics has been forthcoming, and only a few models for a primordial origin have been proposed \cite{Gordon07, DDR07, EKC08}.

In Ref. \cite{EKC08}, Erickcek, Kamionkowski, and Carroll analyzed how a superhorizon fluctuation in an inflationary field could generate such a power asymmetry.  We found that the power asymmetry cannot be reconciled with single-field slow-roll inflation; the superhorizon fluctuation in the inflaton field that is required to generate the observed asymmetry would also induce unacceptable anisotropy in the CMB temperature on large angular scales.  We then considered an alternative inflationary theory, the curvaton model \cite{Mollerach90, LM97, LW02, MT01}, which had been suggested as a possible source of a power asymmetry \cite{LM06, Gordon07}.  In the curvaton model, the inflaton field dominates the Universe's energy density during inflation and drives the inflationary expansion, but the primordial fluctuations arise from quantum fluctuations in a subdominant scalar field called the curvaton.  In Ref. \cite{EKC08}, we showed that a superhorizon fluctuation in the curvaton field can generate the observed asymmetry while respecting both the homogeneity constraints imposed by the CMB \cite{ECK08} and the constraints imposed by upper limits to non-Gaussianity \cite{YW08, WMAP5, SHSHP08, SSZ09, CMB09}.

The model discussed in Ref. \cite{EKC08} predicts that the magnitude and direction of the power asymmetry are scale-invariant.  There are indications, however, that the asymmetry in the CMB temperature fluctuations has a smaller amplitude at $\ell\simeq 220$ \cite{DD05} and does not extend to $\ell \gsim 600$ \cite{Lew08, HBGEL08}.  Furthermore, an analysis of quasar number counts reveals that any asymmetry in the direction $(\ell, b) = (225^\circ, -27^\circ)$ in the rms amplitude of primordial density fluctuations on scales that form quasars ($k \simeq 1.3h-1.8h$ Mpc$^{-1}$) must correspond to $A\lsim0.012$ at the 95\% C.L., assuming that the perturbations are adiabatic \cite{Hirata09}.  In this article, we consider how a superhorizon fluctuation in the curvaton field could produce a scale-dependent power asymmetry that is more pronounced on large scales than on small scales.  

It is possible to dilute the power asymmetry on smaller scales by introducing discontinuities in the inflaton potential and its derivative that change the relative contributions of the curvaton and inflaton fields to the primordial perturbations \cite{Gordon07, EKC08}.  We examine this broken-scale-invariance model in Appendix \ref{app:varyingXi} and find that the discontinuity in the inflaton potential required to satisfy the quasar constraint on the asymmetry violates constraints from ringing in the power spectrum \cite{Covi06, HCMS07}.  In Appendix \ref{app:varyingXi} we also find that it is not possible to sufficiently dilute the asymmetry on small scales by smoothly changing the relative contributions of the curvaton and inflaton fluctuations to the primordial power spectrum.

In light of these difficulties, we turn our attention to the dark-matter isocurvature perturbations generated by some curvaton scenarios \cite{MT01, MT02, LUW03, LW03, GMW04, FRV04, LM07, LVW08, KNSST09}.  In the presence of a superhorizon fluctuation in the curvaton field, the power in these isocurvature perturbations will be asymmetric.  Since isocurvature perturbations decay once they enter the horizon, they will contribute more to the large-scale ($\ell \lsim 100$) CMB anisotropies than to the smaller scales probed by quasars.  Even though the asymmetry in the adiabatic perturbations, which is diluted by the inflaton's contribution, and the asymmetry in the isocurvature perturbations are scale-invariant, the total asymmetry will be suppressed on subhorizon scales as the isocurvature perturbations' contribution to the total power decreases.  Consequently, the desired scale-dependence of the asymmetry is a natural feature of isocurvature perturbations.  In this article we demonstrate that, in certain curvaton scenarios that produce dark-matter isocurvature perturbations,
a superhorizon fluctuation in the curvaton field can produce the observed asymmetry in the CMB without violating any other observational constraints.

We begin by briefly reviewing how isocurvature perturbations are generated in the curvaton scenario in Section \ref{sec:isocurvature}, and we review the CMB signatures of isocurvature perturbations in Section \ref{sec:cmb}.  In Section \ref{sec:asymmetry}, we examine how a hemispherical power asymmetry could be created by a superhorizon fluctuation in the curvaton field in two limiting cases of the curvaton scenario.  We find in Section \ref{sec:case1} that it is not possible to generate the observed hemispherical power asymmetry if the curvaton decay created the dark matter because the necessary superhorizon isocurvature fluctuation induces an unacceptably large temperature dipole in the CMB.  In Section \ref{sec:case2}, we show that the observed asymmetry can be generated by a superhorizon curvaton fluctuation if the curvaton's contribution to the dark matter is negligible.  Our model predicts that the asymmetry will have a specific spectrum and that the current bounds on the contribution of isocurvature perturbations to the CMB power spectrum are nearly saturated.  We summarize our findings and discuss these future tests of our model in Section \ref{sec:conclusions}.  As previously discussed, we show in Appendix \ref{app:varyingXi} that it is not possible to give the asymmetry the required scale-dependence by changing the relative contributions of the curvaton and inflaton to the adiabatic power spectrum.  Finally, we provide a more detailed description of how the curvaton isocurvature perturbation can generate a dark-matter isocurvature fluctuation in Appendix \ref{app:isocurv}.

%%%%%%%%%%%%%%%
\section{Isocurvature Perturbations in the Curvaton Scenario}
\label{sec:isocurvature}
%%%%%%%%%%%%%%%%%%
In the curvaton scenario \cite{Mollerach90, LM97, LW02, MT01}, there is a second scalar field present during inflation, and the energy density of this curvaton field is negligible compared to the energy density of the inflaton.  The curvaton $(\sigma)$ is assumed to be a spectator field during inflation; it remains fixed at its initial value $\sigma_*$ and its energy is given by its potential $V(\sigma)=(1/2)m_\sigma^2\sigma^2$, with $m_\sigma \ll H_\mathrm{inf}$, where $H_\mathrm{inf}$ is the Hubble parameter during inflation.  When $H\simeq m_\sigma$ after inflation, the curvaton field begins to oscillate in its potential well, and it behaves like a pressureless fluid until it decays.  We will assume that the curvaton field is non-interacting prior to its decay.

During inflation, quantum fluctuations in the curvaton field [$(\delta\sigma)_\mathrm{rms} = H_\mathrm{inf}/(2\pi)$] generate a nearly scale-invariant spectrum of isocurvature fluctuations.  After the inflaton decays into radiation, the growth of the curvaton energy density relative to the radiation density creates adiabatic perturbations from these isocurvature fluctuations.  If the curvaton decays before any particle species decouples from radiation, then the isocurvature fluctuation is erased after the curvaton decays because isocurvature fluctuations between interacting fluids in thermal equilibrium decay quickly \cite{Weinberg04,LM07}.  If the curvaton decays after a particle species decouples from the radiation, however, there is a lasting isocurvature fluctuation between that species and the radiation in addition to the adiabatic perturbation generated by the growth of the curvaton energy density relative to the radiation density after inflation \cite{MT01, MT02, LUW03, LW03, GMW04, FRV04, LM07, LVW08, KNSST09}.  

We will restrict our attention to scenarios in which the curvaton decays after dark matter freeze-out but prior to the decoupling of any other particle species.  In this case, an isocurvature fluctuation between dark matter and radiation is created.  (We will neglect baryon isocurvature modes, which may arise due to the annihilations of baryons and antibaryons created during curvaton decay \cite{LMP08, MT09}).  In this section, we will summarize how the final adiabatic perturbation and the dark-matter isocurvature fluctuation relate to the initial curvaton perturbation.  A more detailed review of the relevant physics is presented in Appendix \ref{app:isocurv}.

Working in conformal Newtonian gauge, we take the perturbed Friedmann-Robertson-Walker (FRW) metric to be
\beq
\drm s^2 = -(1+2\Psi)\drm t^2 + a^2(t)\delta_{ij}(1-2\Phi)\drm x^i \drm x^j,
\label{metric}
\eeq
where $a$ is normalized to equal one today.   We define
\beq 
\zeta_i \equiv -\Psi - H \frac{\delta \rho_i}{\dot{\rho_i}}
\eeq 
to be the curvature perturbation on surfaces of uniform $i$-fluid density, and 
\beq
\zeta \equiv -\Psi - H \frac{\delta \rho}{\dot{\rho}}= \sum_i \frac{\dot\rho_i}{\dot\rho}\zeta_i
\label{zetadef}
\eeq
is the curvature perturbation on surfaces of uniform total density.  Throughout this paper, a dot refers to differentiation with respect to proper time $t$.  We use the notation $S_{i\gamma} \equiv 3 (\zeta_i - \zr)$, where a subscript $\gamma$ refers to radiation, to describe isocurvature fluctuations.  For any non-interacting fluid, $\zeta_i$ is conserved on superhorizon scales.  In the absence of isocurvature perturbations, $\zeta$ is constant on superhorizon scales, but if there is an isocurvature perturbation, then $\zeta$ evolves due to the changing value of ${\dot\rho_i}/{\dot\rho}$.

Immediately after inflation, there are superhorizon adiabatic fluctuations from inhomogeneities in the inflaton field $\zeta^{(\mathrm{i})}\simeq\zr^{(\mathrm{i})}$ and superhorizon isocurvature fluctuations in the curvaton field given by $\ssigr$.  After curvaton decay, there are superhorizon adiabatic perturbations $\zeta^{(\mathrm{f})}$ and superhorizon dark-matter isocurvature perturbations $\smr$.  These perturbations are related through a transfer matrix:
\beq
\left(\begin{array}{c} \zeta^{(\mathrm{f})} \\ \smr  \end{array} \right) = \left( \begin{array}{cc} 1 & 
\trs \\ 0 & \tss \end{array} \right) \left(\begin{array}{c} \zeta^{(\mathrm{i})} \\ \ssigr  \end{array}\right).
\label{matrix}
\eeq
This transfer matrix is completely general and applicable to the evolution of any mixture of isocurvature and adiabatic perturbations.  The left column indicates that superhorizon adiabatic perturbations do not evolve in the absence of isocurvature fluctuations and that they are incapable of generating isocurvature fluctuations.  The expressions for $\trs$ and $\tss$ are model dependent.

In the curvaton scenario, $\trs$ depends on the fraction of the Universe's energy that is contained in the curvaton field just prior to its decay.  We define
\beq
R\equiv \left[ \frac{3\Omega_\sigma}{4\Omega_\gamma+3\Omega_\sigma+3\Omega_\mathrm{cdm}} \right]^{(\mathrm{bd})},
\eeq
where $\Omega_\gamma$, $\Omega_\sigma$, and $\Omega_\mathrm{cdm}$ are the radiation energy density, curvaton energy density and cold-dark-matter density, respectively, divided by the critical density.  Throughout this paper, quantities with a ``bd" superscript are to be evaluated just prior to curvaton decay.  We will assume that $R\ll1$ so that the curvaton never dominates the energy density of the Universe.   In the limit of instantaneous curvaton decay with $R\ll1$ \cite{LW02}, 
\beq
\trs \simeq \frac{R}{3} \simeq \frac{1}{4}\Omega_\sigma^\mathrm{(bd)}.
\label{trs}
\eeq
A numerical study of curvaton decay in the absence of dark matter and perturbations from the inflaton [$\zeta^{(\mathrm{i})}=0$]  indicates that this instant-decay expression for $\trs$ is accurate to within 10\% provided that $R$ is evaluated when $H = \Gamma_\sigma/1.4$, where $\Gamma_\sigma$ is the curvaton decay rate \cite{MWU03}.

If the dark matter freezes out prior to curvaton decay, then a dark-matter isocurvature perturbation is created when the dark matter freezes out and when the curvaton decays.  For $R \ll1$ and instantaneous curvaton decay \cite{LM07},
\beqa
\tss &=& \left[\frac{(\alpha - 3)\Omega_\sigma^{(\mathrm{fr})}}{2(\alpha - 2) + \Omega_\sigma^{(\mathrm{fr})}}\right]\frac{\Omega_\mathrm{cdm}^\mathrm{(bd)}}{\Omega_\mathrm{cdm}^\mathrm{(bd)}+B_\mathrm{m}\Omega_\sigma^\mathrm{(bd)}}\nonumber \\
&&+\frac{B_\mathrm{m}\Omega_\sigma^\mathrm{(bd)}}{\Omega_\mathrm{cdm}^\mathrm{(bd)}+B_\mathrm{m}\Omega_\sigma^\mathrm{(bd)}}-R,
\label{tss}
\eeqa
where quantities with an ``fr" superscript are to be evaluated when the dark matter freezes out.  In this expression,  $\alpha \equiv \left.{\frac{\drm \ln \Gamma_\mathrm{cdm}}{\drm \ln T}}\right|_{(\mathrm{fr})}$ gives the dependence of the rate for dark matter annihilations $\Gamma_\mathrm{cdm}$ on temperature $T$ for s-wave annihilations ($\alpha \simeq 21$ for neutralino dark matter), and $B_\mathrm{m} \equiv \Gamma_{\sigma\rightarrow\mathrm{m}}/\Gamma_\sigma$ is the fraction of the curvaton energy that is turned into dark matter when the curvaton decays.  Eq.~(\ref{tss}) differs slightly from the expression for $\tss$ in Ref. \cite{LM07}, but the two expressions are equivalent because $\Omega_\sigma^{(\mathrm{fr})}/\Omega_\sigma^{(\mathrm{bd})}=\Omega_\mathrm{cdm}^{(\mathrm{fr})}/\Omega_\mathrm{cdm}^{(\mathrm{bd})}$.  Numerical studies confirm that this expression for $\tss$ is accurate provided that the decay of the curvaton does not trigger a second era of dark matter self-annihilation \cite{LM07}, as discussed in Appendix \ref{app:isocurv}.

It will be useful to make the $R$ dependence of $\tss$ explicit by defining 
$\tilB R \equiv B_\mathrm{m} \Omega_\sigma^\mathrm{(bd)}/ \Omega_\mathrm{cdm}^\mathrm{(bd)}$
to be the dark matter density from curvaton decay divided by the dark matter density prior to curvaton decay.  We will also define $\tilde{\lambda} \equiv (4/3) \sqrt{{H^{(\mathrm{bd})}}/{H^{(\mathrm{fr})}}}$ so that $\Omega_\sigma^{(\mathrm{fr})} = \tilde{\lambda}R$ (see Appendix \ref{app:isocurv}).  In this notation,
\beq
\tss = \left[ \frac{(\alpha - 3) \tilde{\lambda}R}{2(\alpha-2)+ \tilde{\lambda}R}\right]\left(\frac{1}{1+\tilB R}\right)+\left(\frac{\tilB R}{1+\tilB R}\right) - R.
\eeq
In our analysis, we will consider two limiting cases: $\tilB R\gg1$ (i.e. the curvaton creates nearly all the dark matter), and $\tilB R\ll 1$ (i.e. the curvaton creates an insignificant fraction of the dark matter).  In both cases, we will still assume that $R\ll1$.

If the curvaton creates nearly all the dark matter so that $\tilB R\gg 1$, then
\beq
\lim_{\tilB R\gg 1} \tss = \left[ \frac{(\alpha - 3) \tilde{\lambda}R}{2(\alpha-2)+ \tilde{\lambda}R}\right]\left(\frac{1}{\tilB R}\right)+1-R.
\label{tss1}
\eeq
When we recall that $\tilde\lambda < 1$ is required to make the dark matter freeze-out prior to curvaton decay, we see that the first term in Eq.~(\ref{tss1}) is proportional to $\tilde\lambda/\tilB$, which is much smaller than $R$ if $\tilB R\gg 1$.  The first term is therefore negligible, and we are left with
\beq
\lim_{\tilB R \gg 1} \tss = 1-R.
\label{tssCase1}
\eeq
In the opposite limit, in which the curvaton's contribution to the dark matter density is negligible, we have
\beq
\lim_{\tilB R\ll 1} \tss = \left[ \frac{(\alpha - 3)\tilde{\lambda}}{2(\alpha-2)} + \tilB -1 \right]R \equiv \kappa  R
\label{tssCase2}
\eeq
The first two terms in Eq.~(\ref{tssCase2}) are positive by definition, so $\kappa\gsim-1$.  The first term is always less than 0.5 since $\tilde\lambda<1$, but $\tilB = (4/3) B_\mathrm{m}/\Omega_\mathrm{cdm}^\mathrm{(bd)}$ could be much larger than unity since $\Omega_\mathrm{cdm}^\mathrm{(bd)}\ll1$.   The only upper limit on $\kappa$ is given by $\tilB R\ll 1$ which implies that $\kappa \ll 1/R$.

%%%%%%%%%%%%%%%%%
\section{Isocurvature Modes in the Cosmic Microwave Background}
\label{sec:cmb}
%%%%%%%%%%%%%%%%%

Now that we have defined $\trs$ and $\tss$ in Eq.~(\ref{matrix}), we can relate the early-time perturbations in the matter-radiation fluid to the inflaton and curvaton perturbations created during inflation.  The power spectra of the early-time perturbations in the matter-radiation fluid ($\zeta^{(\mathrm{f})}$ and $\smr$) are the spectra that we will use as initial conditions to calculate the CMB power spectrum.  

Following Ref. \cite{KMV05}, we define
\beqa
{\cal P}_\zeta(k) & \equiv & \frac{k^3}{2\pi^2} \langle \zf(\vec{k}){\zf}^*(\vec{k}) \rangle,\\
{\cal P}_S(k) & \equiv & \frac{k^3}{2\pi^2} \langle \smr(\vec{k})\smr^*(\vec{k}) \rangle,\\
{\cal C}_{\zeta S}(k) &\equiv &\frac{k^3}{2\pi^2} \langle \zf(\vec{k})\smr^*(\vec{k}) \rangle.
\eeqa
We will use a similar convention for the perturbations from inflation:
\beqa
{\cal A}^2\left(\frac{k}{k_0}\right)^{n_\phi - 1} & \equiv & \frac{k^3}{2\pi^2} \langle \zi(\vec{k}){\zi}^*(\vec{k}) \rangle,\\
{\cal B}^2 \left(\frac{k}{k_0}\right)^{n_\sigma - 1} & \equiv & \frac{k^3}{2\pi^2} \langle \ssigr(\vec{k})\ssigr^*(\vec{k}) \rangle.
\eeqa
Both spectra produced during inflation are nearly flat (e.g. \cite{LW02}), and we will assume that $n_\phi\simeq n_\sigma \simeq 1$.   The initial curvature fluctuations are created by the inflaton; the standard slow-roll power spectrum is
\beq
{\cal A}^2 = \frac{G H_\mathrm{inf}^2}{\pi \eh}
\label{Asqrd}
\eeq
where $\eh \equiv  -\dot{H}_\mathrm{inf}/H_\mathrm{inf}^2$ is a slow-roll parameter.  When both the radiation from inflaton decay and the curvaton field are perturbed, $\ssigr \simeq 2\delta\sigma_*/\bar\sigma_*$, where $\delta\sigma_*$ and $\bar\sigma_*$ are evaluated at horizon exit \cite{LV04}.  For superhorizon fluctuations, $\delta\sigma$ and $\bar\sigma$ obey the same evolution equation, so the ratio $\delta\sigma/\bar\sigma$ is conserved \cite{LW03}.  Given that ${\cal P}_{\delta\sigma} = [H_\mathrm{inf}/(2\pi)]^2$, we have
\beq
{\cal B}^2 = \frac{H_\mathrm{inf}^2}{\pi^2 \bar\sigma_*^2}.
\label{Bsqrd}
\eeq
Since $\zi$ is determined by the inflaton fluctuation and $\ssigr$ is determined by the curvaton fluctuation, $\ssigr$ and $\zi$ are uncorrelated.  From Eq.~(\ref{matrix}), we see that
\beqa
{\cal P}_{\zeta}(k) &=& {\cal A}^2 + \trs^2 {\cal B}^2, \\
{\cal P}_{S}(k) &=& \tss^2 {\cal B}^2,\\
{\cal C}_{\zeta S}(k) &=& \trs\tss {\cal B}^2.
\eeqa

\begin{figure}
\centerline{\epsfig{file=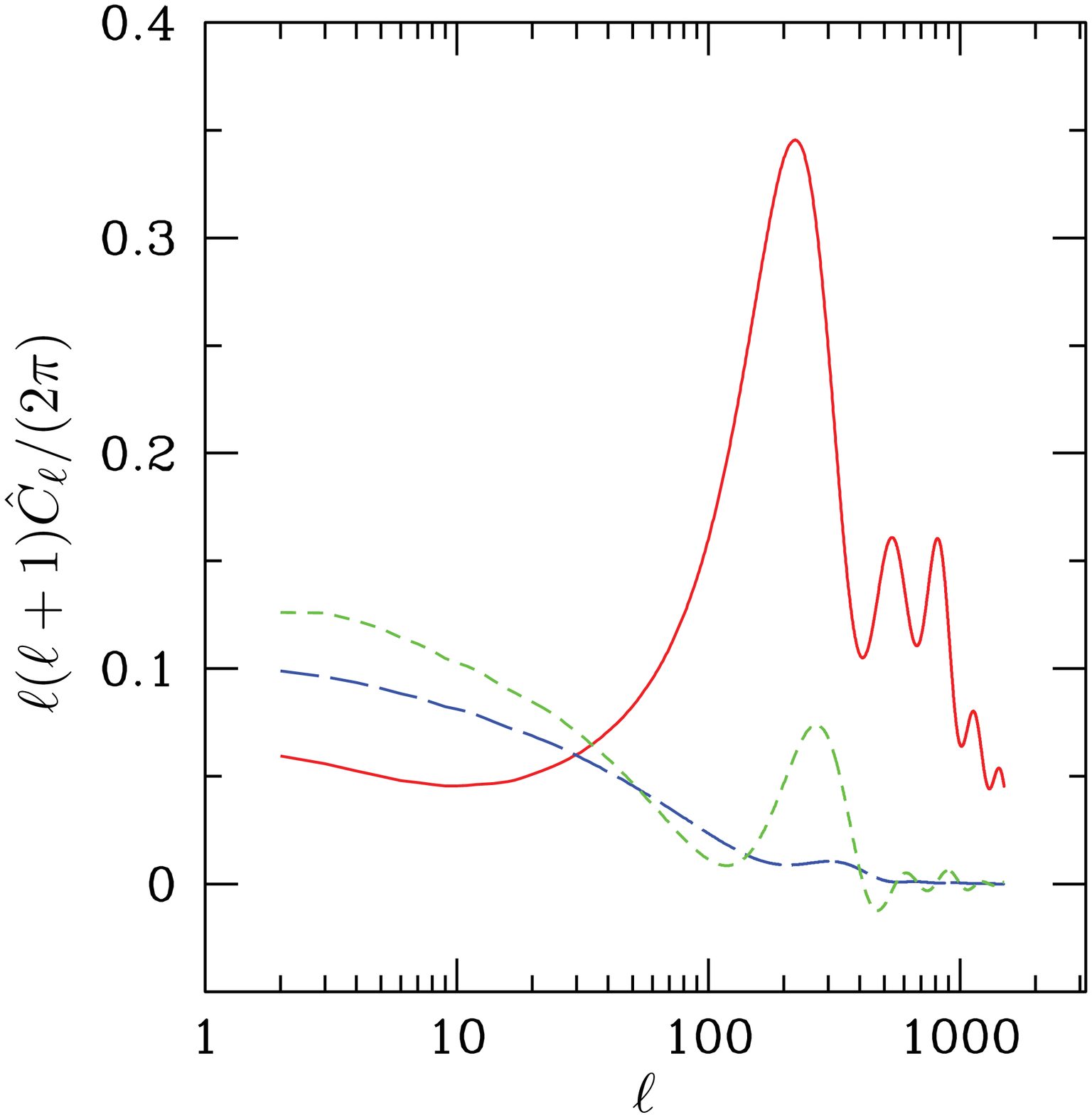, width=3.25in}}
\caption{CMB power spectra for unit-amplitude initial perturbations.  The solid red curve is $\Cad$: the power spectrum derived from ${\cal P}_\zeta(k)=1$.  The long-dashed blue curve is $\Ciso$: the power spectrum derived from ${\cal P}_S(k)=1$.  The short-dashed green curve is $\Ccor$: the difference between the power spectrum derived from ${\cal P}_\zeta(k)={\cal P}_S(k)={\cal C}_{\zeta S}(k)=1$ and $\Cad+\Ciso$.}
\label{fig:Chat}
\end{figure}

The CMB power spectrum may be divided into contributions from adiabatic and isocurvature perturbations \cite{KMV05}:
\beq
C_\ell = \left(\Asqrd+\trs^2\Bsqrd\right)\Cad+{\tss^2\Bsqrd}\Ciso + \trs\tss\Bsqrd\Ccor.
\label{totalCl}
\eeq
In this decomposition, $\Cad$ is the CMB power spectrum derived from a flat spectrum of adiabatic fluctuations with ${\cal P}_\zeta(k)=1$, and $\Ciso$ is the CMB power spectrum derived from a flat spectrum of dark-matter isocurvature perturbations with ${\cal P}_S(k)=1$.  If both isocurvature and adiabatic perturbations are present, with ${\cal P}_\zeta(k)={\cal P}_S(k)={\cal C}_{\zeta S}(k)=1$, then the CMB power spectrum is $\Cad+\Ciso+\Ccor$.  Figure \ref{fig:Chat} shows these three component spectra, as calculated by CMBFast \cite{SZ96} with WMAP5 best-fit cosmological parameters \cite{WMAP5}: $\Omega_\mathrm{b}=0.0462, \,\Omega_\mathrm{cdm}=0.233, \,\Omega_\Lambda=0.721$ and $H_0=70.1$ km/s/Mpc.

Figure \ref{fig:Chat} clearly shows that isocurvature perturbations leave a distinctive imprint on the CMB power spectrum.  It is therefore possible to constrain the properties of ${\cal P}_{\zeta}(k)$ and ${\cal P}_{S}(k)$ using CMB data.  These constraints are often reported as bounds on the isocurvature fraction $\alpha$ and the correlation parameter $\gamma$:
\beqa
\alpha &\equiv& \frac{\tss^2\Bsqrd}{\Asqrd + \trs^2\Bsqrd+\tss^2\Bsqrd}, \label{alpha}\\
\gamma &\equiv& \mathrm{sign}(\trs\tss)\frac{\trs^2\Bsqrd}{\Asqrd + \trs^2\Bsqrd}.
\eeqa
We will find it useful to define $\xi$ to be the fraction of adiabatic perturbations from the curvaton:
\beq
\xi \equiv \frac{\trs^2\Bsqrd}{\Asqrd + \trs^2\Bsqrd},
\eeq
with $\trs = R/3$.  We then see that
\beqa
\alpha &=& \frac{9(\xi/R^2)\tss^2}{1+9(\xi/R^2)\tss^2} \label{alpha2}\\
\gamma &=& \mathrm{sign}(\tss)\xi.
\eeqa

Ideally, we would like to use constraints for $\alpha$ and $\gamma$ that were derived assuming only that $n_\mathrm{ad} \simeq n_\mathrm{iso} \simeq1$, where $n_\mathrm{ad}$ and $n_\mathrm{iso}$ are the spectral indices for ${\cal P}_\zeta(k)$ and ${\cal P}_S(k)$ respectively.   Such an analysis does not exist, although constraints have been derived for the $n_\mathrm{ad}=n_\mathrm{iso}$ case and have found that $n_\mathrm{ad}=n_\mathrm{iso}\simeq 1$ gives the best fit to observations \cite{BDP06, SCH09}.  Using WMAP3 data and large-scale structure data, Ref. \cite{BDP06} found that $\alpha<0.15$ at 95\% confidence, with a slight preference for negative values of $\gamma$, although $\gamma=0$ was included in the 1$\sigma$ interval.  Ref. \cite{SCH09} updated this analysis and found similar constraints on $\gamma$, but unfortunately they did not report a constraint on $\alpha$.
Meanwhile, the most general analyses \cite{KMV05, BGLV05, BDP06, KKMV07, SCH09} make no assumptions regarding $n_\mathrm{iso}$ and conclude that models with $n_\mathrm{iso} \simeq 2-4$ provide the best fit to the data.  Since their bounds on $\alpha$ and $\gamma$ are marginalized over $n_\mathrm{iso}$ values that are unreachable in the curvaton scenario, these constraints are not applicable to our model.  

There are analyses that specifically target the curvaton scenario, but they assume that the curvaton generates all of the primordial fluctuations (i. e. $\Asqrd \ll \trs^2\Bsqrd$) \cite{BGLV05, BDP06, WMAP5}.  In this case, $\xi = 1$, and the isocurvature and adiabatic fluctuations are completely correlated or anti-correlated, depending on the sign of $\tss$.   Furthermore, Eq.~(\ref{alpha2}) shows that $\alpha \gsim 0.9$ if $\xi = 1$ and $\tss^2 \gsim R^2$.  Since this high value for $\alpha$ is thoroughly ruled out, these analyses of isocurvature perturbations in the curvaton scenario disregard the possibility that $\tilB R\ll 1$ and assume that most of the dark matter is created by curvaton decay.  In this case, $\tss$ is given by Eq.~(\ref{tssCase1}) and the derived upper bound on $\alpha$ ($\alpha <0.0041$ from Ref. \cite{WMAP5}) implies that $R>0.98$.  Since we require $R\ll1$, we can conclude that we will be restricted to mixed-perturbation scenarios in which both the curvaton and the inflaton contribute to the adiabatic perturbation spectrum and $\xi <1$.

Finally, some analyses constrain completely uncorrelated ($\gamma = 0$) isocurvature and adiabatic perturbations (a.k.a. axion-type isocurvature) with $n_\mathrm{iso}=1$ \cite{BGL07, WMAP5, HHRW09}.   These constraints are most relevant to our models, however, because we will see that $\xi=|\gamma|$ must be small to create an asymmetry that vanishes on small scales.  (The discussion in the previous paragraph also foreshadows the fact that $\xi \ll 1$ will be necessary to obtain $R\ll1$.)  Furthermore, we will show that only models with $\gamma>0$ can generate the observed asymmetry, so the constraints derived in Refs. \cite{BDP06, SCH09} are too generous.  We will therefore use the bound on $\alpha$ derived for uncorrelated adiabatic and isocurvature perturbations in our analysis.  WMAP5 data alone constrains $\alpha<0.16$ at 95\% confidence, but the upper bound on $\alpha$ is significantly reduced if BAO and SN data are used to break a degeneracy between $\alpha$ and $n_\mathrm{ad}$.  With the combined WMAP5+BAO+SN dataset, the 95\% C.L. upper bound on $\alpha$ is \cite{WMAP5}
\beq
\alpha < 0.072,
\eeq
with a best-fit value of $n_\mathrm{ad}\simeq1$.   Other analyses have found similar bounds: $\alpha<0.08$ \cite{BGL07} and $\alpha<0.09$ \cite{HHRW09} at 95\% C.L. 

The other observable effect of isocurvature fluctuations that we must consider is non-Gaussianity \cite{KNSST08, LVW08, HKMTY09, KNSST09, KKNT09}.  Following Ref. \cite{HKMTY09}, we define $f_\mathrm{NL}^{(\mathrm{iso})}$ through
\beq
\smr = \eta + f_\mathrm{NL}^{(\mathrm{iso})}\left(\eta^2-\langle{\eta^2}\rangle\right),
\label{fnliso_def}
\eeq
where $\eta$ is drawn from a Gaussian probability spectrum.$^1$\footnotetext[1]{This definition of $f_\mathrm{NL}^{(\mathrm{iso})}$ differs from the definition given in Refs. \cite{KNSST08, KNSST09, KKNT09}, which define $f_\mathrm{NL}^{(\mathrm{iso})}$ in terms of the bispectrum of curvature perturbations during matter domination.  That $f_\mathrm{NL}^{(\mathrm{iso})}$ includes information about the isocurvature fraction $\alpha$ and is consequently much smaller than the $f_\mathrm{NL}^{(\mathrm{iso})}$ defined in Eq.~(\ref{fnliso_def}) for a given model.} This is analogous to the definition of $f_\mathrm{NL}$ in terms of the gravitational potential for adiabatic perturbations \cite{VWHK00,KS01}.  
Given the current upper bound on $\alpha$, $f_\mathrm{NL}^{(\mathrm{iso})}\simeq 10^4$ produces a CMB bispectrum that is equal in magnitude to the CMB bispectrum if $f_\mathrm{NL} \lsim 17$ for purely adiabatic perturbations \cite{HKMTY09}.  Furthermore, non-Gaussianity from isocurvature perturbations is distinguishable from adiabatic non-Gaussianity through the scale dependence of the bispectrum \cite{KNSST08, KNSST09, HKMTY09}, and an analysis of the WMAP5 data with Minkowski functionals found $-2.5 \times 10^4 <f_\mathrm{NL}^{(\mathrm{iso})}<2.0\times10^4$ at 95\% C.L. for $\alpha =0.072$ \cite{HKMTY09}.

For isocurvature perturbations from the curvaton,
\beq
\smr=\tss\ssigr=\tss\left[2\frac{\delta\sigma_*}{\bar\sigma_*}+\left(\frac{\delta\sigma_*}{\bar\sigma_*}\right)^2\right],
\label{smr}
\eeq
and we can set $\eta = 2\tss{\delta\sigma_*}/{\bar\sigma_*}$.  Thus we see that 
\beq
f_\mathrm{NL}^{(\mathrm{iso})} = \frac{1}{4\tss}
\label{fnliso}
\eeq
for the curvaton model.  Given that $|f_\mathrm{NL}^{(\mathrm{iso})}|\lsim2.5\times10^4$ for $\alpha=0.072$, we see that $|\tss|\gsim  10^{-5}$ is required if the current bound on isocurvature power is saturated.  The curvaton also introduces non-Gaussianity in the adiabatic perturbations; since the fluctuations from the inflaton are Gaussian \cite{Maldacena03}, $f_\mathrm{NL}$ for mixed perturbations from the inflaton and curvaton is given by \cite{LUW03, ISTY08}
\beq
f_\mathrm{NL} = \frac{5\xi^2}{4R}.
\label{fnlad}
\eeq
The current upper limit on $f_\mathrm{NL}$ from the CMB and large-scale structure is $f_\mathrm{NL} \lsim 80$ \cite{YW08, WMAP5, SHSHP08, SSZ09, CMB09}.

%%%%%%%%%%%%%%%%%
\section{A Power Asymmetry from Curvaton Isocurvature}
\label{sec:asymmetry}
%%%%%%%%%%%%%%%%%

\begin{figure}[b]
\includegraphics[width=8.5cm]{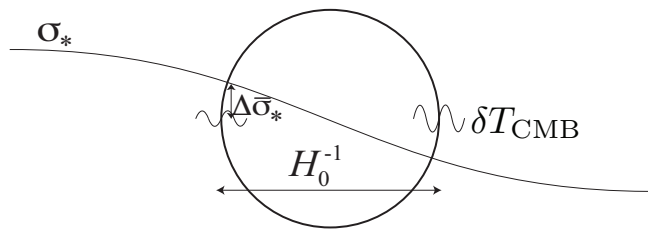}
\caption{Measurements of temperature fluctuations in the CMB show that the rms temperature-fluctuation amplitude is larger in one side of the sky than in the other.  We propose that this asymmetry is generated by a large-amplitude  fluctuation in the initial value of the curvaton field $\sigma_*$.  The fluctuation in $\sigma_*$ across the observable Universe is $\Delta\bar\sigma_*$.}
\label{fig:supermode}
\end{figure}

In an earlier article \cite{EKC08}, Erickcek, Kamionkowski, and Carroll proposed that the hemispherical power asymmetry in the CMB could result from a large-amplitude superhorizon fluctuation in the curvaton field $\sigma$, as depicted in Fig.~\ref{fig:supermode}.  The difference between $\bar\sigma_*$ on one side of the surface of last scatter and its average value in the observable Universe, $\Delta\bar\sigma_*$, will introduce a power asymmetry $\Delta C_\ell$ in the CMB through Eq.~(\ref{totalCl}).  The CMB power spectrum depends on $\bar\sigma_*$ through $\Bsqrd$, as given by Eq.~(\ref{Bsqrd}), and through $\trs$ and $\tss$, which are functions of $R$.  For $R\ll1$, the Universe is radiation-dominated between the end of inflation and the decay of the curvaton, and 
\beq
R = \pi \left(\frac{\bar\sigma_*}{\mpl}\right)^2 \sqrt{\frac{1.4\,m_\sigma}{\Gamma_\sigma}},
\eeq
where $\mpl^2 \equiv G^{-1}$ is the Planck mass \cite{LW02}.  Differentiating Eq.~(\ref{totalCl}) with respect to $\sigma_*$ gives
\beqa
\Delta C_\ell &=& 2 \frac{\Delta\bar\sigma_*}{\bar\sigma_*}\Bsqrd\left[{\frac{R^2}{9}\Cad-\left(\tss^2-2\tss R \frac{\drm \tss}{\drm R}\right) \Ciso } \right. \nonumber \\
&&\left.\quad\quad\quad\quad\,\,+ \,\frac{R^2}{3}\frac{\drm \tss}{\drm R}\Ccor\right]
\label{delCgen}
\eeqa
where we have used $\trs = R/3$ for the curvaton scenario.  

In Ref. \cite{EKC08}, we assumed that the curvaton decayed prior to dark matter freeze-out so that no isocurvature perturbations are created.  In this scenario, $\tss=0$, and 
\beq
\frac{\Delta C_\ell}{C_\ell} = 2\frac{\Delta\bar\sigma_*}{\bar\sigma_*}\xi.
\eeq
This power asymmetry is scale-invariant.  However, if the curvaton also generates isocurvature perturbations, the power asymmetry will be scale-dependent due to the differences between $\Cad, \Ciso$, and $\Ccor$ shown in Fig.~\ref{fig:Chat}.  We will extract this scale-dependence by defining $K_\ell$ through
\beq
\left|\frac{\Delta C_\ell}{C_\ell} \right| \equiv 2\frac{\Delta\bar\sigma_*}{\bar\sigma_*}K_\ell.
\eeq

The dipolar modulation parameter $A$ used by Refs.~\cite{Eriksen07, Eriksen09} and defined in Eq.~(\ref{defA}) describes the asymmetry in the amplitude of temperature fluctuations, so for small $A$, $A \simeq (1/2)(\Delta C_\ell/C_\ell)$.  The modulation is assumed to be scale-invariant and is measured for $\ell \leq \ell_\mathrm{max}$.  To relate the scale-dependent power asymmetry described by $K_\ell$ to $A$, we assume that all modes with $2\leq\ell \leq \ell_\mathrm{max}$ are weighted equally in determining the measured asymmetry.
Since there are $ (\ell_\mathrm{max}-1)(\ell_\mathrm{max}+3)$ modes in total,
\beq
A = \frac{\Delta\bar\sigma_*}{\bar\sigma_*}\sum_{\ell=2}^{\ell_\mathrm{max}} \frac{2\ell+1}{ (\ell_\mathrm{max}-1)(\ell_\mathrm{max}+3)} K_\ell \equiv  \frac{\Delta\bar\sigma_*}{\bar\sigma_*} \Atil.
\label{Adef}
\eeq
We note that $\Atil$ does not depend on the amplitude of the superhorizon fluctuation; it is determined by $\tss$, $R$, and $\xi$.  Since $\Delta\bar\sigma_*$ cannot be larger than $\bar\sigma_*$, $\Atil$ is the largest asymmetry that can be produced by a particular curvaton scenario.  Unless otherwise noted, we set $\ell_\mathrm{max} = 64$ to match Ref. \cite{Eriksen09}.  As mentioned previously, Ref. \cite{Eriksen09} found that $A = 0.072 \pm 0.022$ for $\ell \lsim 64$, yet the isotropic distribution of quasars constrains $A\lsim0.012$ for $k \simeq 1.3h-1.8h$ Mpc$^{-1}$ \cite{Hirata09}.

In the following subsections we will examine $K_\ell$ for the two limiting cases discussed in Section \ref{sec:isocurvature}.  First, we will consider scenarios in which most of the dark matter is created during curvaton decay and $\tss \simeq1-R$.  Then we will consider scenarios in which the curvaton's contribution to the dark matter is negligible and $\tss = \kappa R$ with $-1\lsim \kappa \lsim 1/R$.  In both cases, we will see that $K_\ell$ decreases rapidly when $\ell\gsim 10$.  We will also find that the superhorizon curvaton fluctuation required to generate the observed asymmetry must have a large amplitude:  $\Delta\bar\sigma_*/\bar\sigma_* \gsim 1/2$.  We therefore must consider how this large-amplitude superhorizon fluctuation will create large-scale anisotropies in the CMB through the Grishchuk-Zel'dovich effect \cite{GZ78}.

A superhorizon adiabatic fluctuation does not induce a prominent temperature dipole in the CMB due to a cancellation between the intrinsic dipole and the Doppler dipole \cite{ECK08, ZS08}, but this is not the case for superhorizon isocurvature fluctuations \cite{LP96, Langlois96}.  After matter domination, the evolution of the potential $\Psi$ and the fluid velocity's dependence on $\Psi$ are the same for adiabatic and isocurvature initial conditions \cite{HS95}.  Therefore, the induced Doppler dipole and the anisotropy from the integrated Sachs-Wolfe effect will be the same for adiabatic and isocurvature fluctuations if the surface of last scatter is taken to be in the matter-dominated era.  The only difference between the CMB dipole induced by an adiabatic perturbation and the CMB dipole induced by an isocurvature perturbation arises from the Sachs-Wolfe anisotropy; for adiabatic perturbations $(\Delta T/T)_\mathrm{SW} = \Psi_\mathrm{dec}/3$, while $(\Delta T/T)_\mathrm{SW} = 2\Psi_\mathrm{dec}$ for isocurvature perturbations, where $\Psi_\mathrm{dec}$ is evaluated at the time of decoupling.  Since we know that the integrated Sachs-Wolfe effect and the Doppler dipole exactly cancel the Sachs-Wolfe anisotropy for adiabatic perturbations, the residual temperature dipole for isocurvature fluctuations must be $5 \Psi_\mathrm{dec}/3$.  

If $S_0$ is the initial matter isocurvature fluctuation set deep in the radiation-dominated era, then $\Psi_\mathrm{dec} = -S_0/5$ \cite{HS95}.  We are considering dark-matter isocurvature fluctuations, so we have $S_0 = \smr\Omega_\mathrm{cdm}/(\Omega_\mathrm{cdm}+\Omega_\mathrm{b})$, where $\smr$ is given by Eq.~(\ref{smr}).  We treat the superhorizon fluctuation in the curvaton field as a sine wave: $\delta \sigma_* = \sigma_{\vec{k}} \sin (\vec{k} \cdot \vec{x})$, where $k\ll H_0$.  By choosing this form for $\delta \sigma$, we have placed ourselves at the node of the sine wave, but the constraints we derive on $\delta \sigma$ are not strongly dependent on this choice \cite{ECK08}.  We decompose the CMB temperature anisotropy into multipole moments,
\beq
\frac{\delta T}{T}(\hat{n}) = \sum_{\ell,m} a_{\ell m} Y_{\ell m}(\hat{n}),
\eeq
and we find that, to lowest order in $kx_\mathrm{dec}$, where $x_\mathrm{dec}$ is the comoving distance to the last scattering surface, the dipolar moment generated by the superhorizon curvaton fluctuation is 
\beq
a_{10 } = -\frac{1}{3}\sqrt{\frac{4\pi}{3}}\left(kx_\mathrm{dec}\right)\frac{\Omega_\mathrm{cdm}}{\Omega_\mathrm{cdm}+\Omega_\mathrm{b}} \left(2\tss\frac{ \sigma_{\vec{k}}}{\bar\sigma_*}\right)
\eeq
where we have chosen axes that are aligned with the asymmetry $(\hat{z}=\hat{k})$.  The variation in $\sigma$ across the surface of last scattering is $\Delta \bar\sigma_* = \sigma_{\vec{k}}(kx_\mathrm{dec})$, and it is constrained by the CMB temperature dipole:
\beq
\tss \left(\frac{\Delta\bar\sigma_*}{\bar\sigma_*}\right) \lsim 0.9 {\cal D},
\label{dipole}
\eeq
where ${\cal D}$ is the largest value of $| a_{10 }|$ that is consistent with observations of the CMB dipole.  The observed temperature dipole has amplitude $\Delta T/T = (1.231 \pm 0.003) \times 10^{-3} $ and it is not aligned with the asymmetry \cite{WMAPdip}.  At least a portion of this anisotropy is attributable to our proper motion, but recent attempts to measure the peculiar velocity of the local group have found that the measured velocity is smaller than the velocity predicted by the CMB and misaligned with the temperature dipole, with a difference of $~500$ km/s  \cite{Erdo06, LTM08}.  We therefore take ${\cal D} = 0.0034$, which corresponds to a velocity of $500$ km/s, to generate a conservative upper bound.

The superhorizon fluctuation in the curvaton field will also generate a quadrupolar anisotropy in the CMB.  The induced quadrupole is higher-order in $\Delta \bar\sigma_*/\bar\sigma_*$ because it originates from the quadratic term in $\ssigr$ [see Eq.~(\ref{smr})].   Nevertheless, the upper bound on $\Delta \bar\sigma_*/\bar\sigma_*$ from the CMB quadrupole is similar to the bound from the dipole because observations of the CMB quadrupole are not contaminated by our proper motion.  The CMB quadrupole is the sum of contributions from the superhorizon isocurvature perturbation and the superhorizon adiabatic perturbation ($\zeta = \trs \ssigr$ during radiation domination) generated by the curvaton field.  In the coordinate system aligned with the asymmetry,
\beqa
a_{20}&=& -\frac{1}{3}\sqrt{\frac{4\pi}{5} }\left(kx_\mathrm{dec}\right)^2\left(\frac{ \sigma_{\vec{k}}}{\bar\sigma_*}\right)^2 \label{fullquad}\\ 
&&\times\left[\delta_2^\mathrm{ad}\left(\frac{2R}{5}\right)+\delta_2^\mathrm{iso}\frac{\Omega_\mathrm{cdm}}{\Omega_\mathrm{cdm}+\Omega_\mathrm{b}}\left(\frac{2\tss}{5}\right)\right], \nonumber
\eeqa
where $\delta_2^\mathrm{ad}$ is derived by analyzing the Sachs-Wolfe effect, the integrated Sachs-Wolfe effect, and the fluid velocity at the surface of last scatter generated by a superhorizon adiabatic perturbation (see Ref. \cite{ECK08}).  In the limit that decoupling occurs after matter-domination,  $\delta_2^\mathrm{ad} = 0.338$, and $\delta_2^\mathrm{iso} = 5/3+\delta_2^\mathrm{ad}$ since only the contribution from the Sachs-Wolfe effect is different for isocurvature initial conditions.  It follows that the upper-bound on $\Delta \bar\sigma_*/\bar\sigma_*$ from the CMB quadrupole is
\beq
\left(0.34\, R+1.67\, \tss \right)\left(\frac{\Delta\bar\sigma_*}{\bar\sigma_*}\right)^2 \lsim 4.7{Q},
\label{quad}
\eeq
where ${Q}$ is the largest value of $| a_{20 }|$ that is consistent with observations of the CMB quadrupole.   As discussed in Refs. \cite{EKC08, ECK08}, contributions to $a_{20}$ from smaller scale perturbations could partially cancel the contribution to $a_{20}$ from a superhorizon perturbation.  The power in these fluctuations is given by the predicted value for $C_2$ in the best-fit \lcdm model, $C_2 =  1.7 \times 10^{-10}$, and the WMAP5 ILC map gives $a_{20} = 7.3 \times 10^{-6}$.  We will focus on models with $\tss>0$, and we see from Eq.~(\ref{fullquad}) that $a_{20}$ is negative in this case. We therefore set $Q =\left|{ 7.3 \times 10^{-6}-2\sqrt{C_2}}\right| = 1.9 \times 10^{-5}$ as a 2$\sigma$ upper bound on the temperature quadrupole induced by the variation in the curvaton field across the observable Universe.

%%%%%%%%%%%%%%%%%
\subsection{Case 1: The curvaton creates most of the dark matter.}
\label{sec:case1}
%%%%%%%%%%%%%%%%%
If most of the dark matter is created when the curvaton decays, then $\tss \simeq 1-R$, as in Eq.~(\ref{tssCase1}).  In this case, Eqs.~(\ref{totalCl}) and (\ref{delCgen}) imply
\beqa
\frac{\Delta C_\ell}{C_\ell} 
&\simeq& 2\frac{\Delta\bar\sigma_*}{\bar\sigma_*}\xi \left[\frac{\Cad-\frac{9}{R^2}\Ciso-3\Ccor}{\Cad + \xi\left(\frac{9}{R^2}\Ciso +\frac{3}{R}\Ccor\right)}\right]
\label{Klexact}
\eeqa
where we have kept only the leading-order term in $R$ in the coefficients of $\Ciso$ and $\Ccor$.  We can also assume that $R\Ccor \ll \Ciso$ since Fig.~\ref{fig:Chat} shows that $\Ciso\simeq\Ccor$.  Finally, if $R\lsim$ 0.01, then $R^2 \Cad \ll \Ciso$ for $\ell \lsim 1500$, and we may neglect $\Cad$ in the numerator.  With these simplifications, we have $\Delta C_\ell /C_\ell = -2 (\Delta\bar\sigma_*/\bar\sigma)K_\ell$ where
\beq
K_\ell \simeq \frac{\frac{9\xi}{R^2}\Ciso}{\Cad+\frac{9\xi}{R^2}\Ciso}.
\label{Klapprox}
\eeq
This approximate expression for $K_\ell$ is useful because it only depends on $\xi/R^2$.  It is accurate to within 1\% for $\ell \leq 1500$ if $R\leq 0.01$ and accurate to within 0.1\% if $R\leq0.001$.  However, it does not have the appropriate limit for $\ell \rightarrow \infty$; since $\Ciso/\Cad \rightarrow 0$ in this limit, Eq.~(\ref{Klexact}) shows that $K_\ell \rightarrow \xi$, but the approximate form goes to zero.

\begin{figure}
\includegraphics[width=8.5cm]{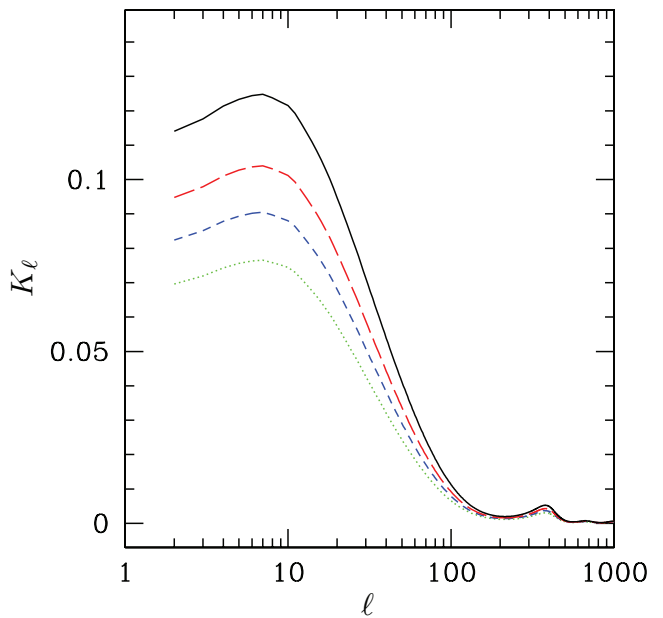}
\caption{$K_\ell$ for scenarios in which most of the  dark matter comes from curvaton decay.  The power asymmetry is given by $\Delta C_\ell /C_\ell = -2 (\Delta\bar\sigma_*/\bar\sigma)K_\ell$.  The solid black curve corresponds to $\xi/R^2 = 0.0086$, which saturates the current bound on power from isocurvature perturbations.  The lower curves have $\xi/R^2 = 0.007$ (long-dashed), 0.006 (short-dashed) and 0.005 (dotted).  For descending values of $\xi/R^2$, these curves correspond to asymmetry amplitudes $\Atil = 0.055, \,0.045,\, 0.039,$ and $0.033$.}
\label{fig:Klcase1}
\end{figure}

Figure \ref{fig:Klcase1} shows the approximate form of $K_\ell$, given by Eq.~(\ref{Klapprox}), for four values of $\xi/R^2$: 0.005, 0.006, 0.007, and 0.0086.  We see that $K_\ell$ increases on large scales as $\xi/R^2$ increases.   On smaller scales, we see that $K_\ell$ is not very sensitive to changes in $\xi/R^2$.  Thus, to obtain the desired asymmetry on large scales and nearly no asymmetry on small scales, we just need to increase $\xi/R^2$!  Unfortunately, the upper bound on the isocurvature fraction $\alpha$ places an upper bound on $\xi/R^2$:
\beq
\alpha < 0.072 \Longrightarrow \frac{\xi}{R^2} < 0.0086.
\eeq
The solid curve in Fig.~\ref{fig:Klcase1} corresponds to $\xi/R^2 = 0.0086$ and is therefore the maximal $K_\ell$ curve that is consistent with the current limits on the isocurvature contribution to the CMB power spectrum.  We also note that satisfying the upper bound on $\alpha$ requires $\xi$ to be much smaller than $R$, and we have assumed that $R\ll1$.  The adiabatic and isocurvature fluctuations are therefore uncorrelated.

Figure \ref{fig:Klcase1} also shows that $K_\ell$ peaks for $\ell\simeq10$ and decreases rapidly as $\ell$ increases from 10 to 100.  Furthermore, the asymmetry nearly vanishes for larger $\ell$, so it will be easy to satisfy the quasar constraint.  The desired scale-dependence comes at a cost though; the smaller values of $K_\ell$ at $\ell\gsim20$ dilute the scale-averaged asymmetry $A$.  For $\xi/R^2 = 0.0086$, the maximal asymmetry is given by $\Atil = 0.055$.  Thus we see that saturating the upper bound on isocurvature power ($\alpha$) and setting $\Delta\bar\sigma_*=\bar\sigma_*$ leads to an asymmetry that is almost $1\sigma$ below the observed value.   Moreover, the curvaton creates most of the dark matter in this scenario; since $\Delta\bar\sigma_*/\bar\sigma_* \simeq 1$ is required to generate sufficient asymmetry, this model requires that the dark matter density varies by a factor of unity across the observable Universe!  Unsurprisingly, such a large isocurvature fluctuation is not consistent with the large-scale homogeneity of the CMB.  Since $\tss\simeq1$ in this scenario, the CMB dipole constrains $\Delta\bar\sigma_*/\bar\sigma_*\lsim 3\times10^{-3}$ from Eq.~(\ref{dipole}).  We conclude that the curvaton cannot generate the observed power asymmetry if the dark matter is created during curvaton decay.

%%%%%%%%%%%%%%%%%
\subsection{Case 2: The curvaton's contribution to the dark matter is negligible.}
\label{sec:case2}
%%%%%%%%%%%%%%%%%

We now turn our attention to the opposite scenario, in which the curvaton's contribution to the dark matter density is insignificant.  In this case, Eq.~(\ref{tssCase2}) tells us that $\tss \simeq \kappa R$, with $-1\lsim\kappa\lsim1/R$.  We will see, however, that this model can generate the observed asymmetry only if $\kappa \lsim 1.4$.  Thus, we will be considering scenarios in which the curvaton generates adiabatic and isocurvature fluctuations that are equal in magnitude ($\tss^2 \simeq \trs^2$), in stark contrast to the scenarios considered in the previous section.  We anticipate that generating comparable adiabatic and isocurvature fluctuations from the curvaton will be advantageous for two reasons.  First, the asymmetry can be partially contained in the adiabatic perturbations, which will make it easier to generate the observed asymmetry without violating the current bounds on isocurvature power.  Second, the superhorizon isocurvature perturbation generated by $\Delta\bar\sigma_*$ will be proportional to $R$ and can therefore be reduced by decreasing $R$.  The downside is that it will be difficult to make the asymmetry sufficiently scale-dependent to satisfy the quasar bound because the adiabatic perturbations are asymmetric as well.

For $\tss = \kappa R$, the power asymmetry generated by the superhorizon curvaton perturbation is given by $\Delta C_\ell /C_\ell = 2 (\Delta\bar\sigma_*/\bar\sigma_*)K_\ell$ where, from Eqs.~(\ref{totalCl}) and (\ref{delCgen}), we have 
\beq
K_\ell = \xi\left[\frac{\Cad + 9\kappa^2\Ciso +3\kappa \Ccor}{\Cad+\xi\left(9\kappa^2\Ciso+3\kappa\Ccor\right)}\right].
\eeq
We see that $K_\ell \rightarrow \xi$ as $\ell\rightarrow\infty$ as expected; on small scales, the only source of asymmetry is the adiabatic power from the curvaton.  We can therefore anticipate that the quasar constraint will place an upper bound on $\xi$.  We also see that all the isocurvature contributions to the power asymmetry are proportional to $\kappa$ or $\kappa^2$, and this implies that the necessary scale-dependence of $K_\ell$ will place a lower limit on $|\kappa|$.

\begin{figure}
\includegraphics[width=8.5cm]{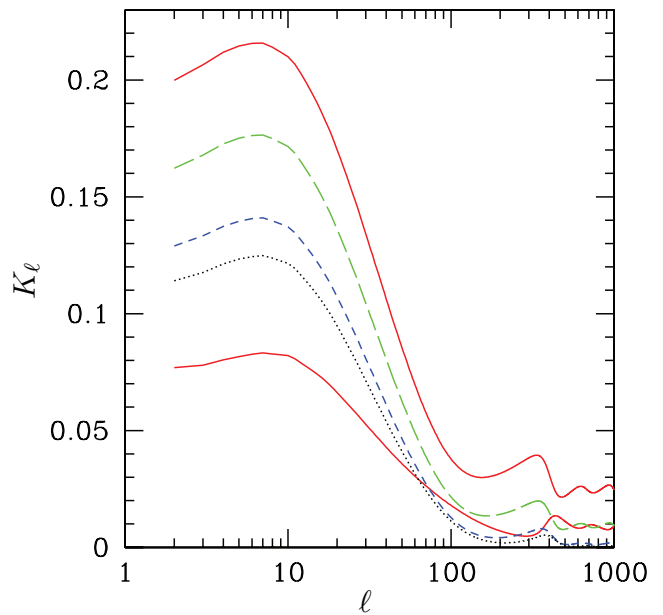}
\caption{$K_\ell$ for scenarios in which the curvaton's contribution to the dark matter density is negligible and $\tss =\kappa R$.  The power asymmetry is given by $\Delta C_\ell /C_\ell = 2 (\Delta\bar\sigma_*/\bar\sigma)K_\ell$.  All of the curves have $\kappa^2\xi = 0.0086$, which saturates the upper limit on isocurvature power.  The top three curves have $\kappa=0.6$ (top, solid), $\kappa =1$ (long-dashed), and $\kappa=3$ (short-dashed).  The maximal scale-averaged asymmetries possible for these curves are $\Atil = 0.11$, $\Atil = 0.081$, and $\Atil = 0.062$.  The dotted curve is the limit as $\kappa\rightarrow\infty$, and it has $\Atil = 0.055$.  The bottom solid curve has $\kappa=-1$ and $\Atil = 0.043$.}
\label{fig:Klcase2}
\end{figure}

Differentiating $K_\ell$ with respect to $\xi$ and $|\kappa|$ reveals that increasing $\xi$ or $|\kappa|$ increases $K_\ell$, unless $-0.2\lsim\kappa\lsim0$, in which case the $\Ciso$ and $\Ccor$ terms partially cancel on large scales, leaving $K_\ell$ nearly scale-invariant.  Unfortunately, the upper limit on isocurvature power places an upper limit on $|\kappa|$ and $\xi$:
\beq
\alpha < 0.072 \Longrightarrow \kappa^2\xi< 0.0086.
\eeq
If we differentiate $K_\ell$ with respect to $|\kappa|$ while keeping $\kappa^2\xi$ fixed, we find that increasing $|\kappa|$ decreases $K_\ell$ for $\kappa > -0.3$ and increases $K_\ell$ for $\kappa < -0.3$.   Furthermore, the $|\kappa|\rightarrow\infty$ limit of $K_\ell$, with fixed $\kappa^2\xi$, is
\beq
\lim_{|\kappa|\rightarrow\infty; \,\,\mathrm{fixed}\,\, \kappa^2\xi} K_\ell = \frac{9\left(\kappa^2\xi\right)\Ciso}{\Cad+9\left(\kappa^2\xi\right)\Ciso}.
\label{Kllimit}
\eeq
Figure \ref{fig:Klcase2} shows $K_\ell$ with $\kappa^2\xi = 0.0086$ for various values of $\kappa$.  We see that as $\kappa$ increases from zero, the curves rapidly approach the dotted curve, which is Eq.~(\ref{Kllimit}).  If we could decrease $\kappa$ toward $-\infty$, the curves would approach this limit from below, but in Section \ref{sec:isocurvature} we found that $\kappa \gsim -1$, which corresponds to the lower solid curve.  

The observed asymmetry is $A=0.072\pm0.022$ for $\ell\lsim64$, which requires $K_\ell \gsim 0.08$ on average over this $\ell$ range.  Comparing Figs. \ref{fig:Klcase1} and \ref{fig:Klcase2} reveals that it is much easier to generate the required asymmetry if the curvaton's contribution to the dark matter is insignificant because the peak in $K_\ell$ is higher for $\tss\simeq R$ than for $\tss \simeq1$.   If we saturate the upper bound on isocurvature power by setting $\kappa^2\xi = 0.0086$, then $\Atil \gsim 0.08$ if $0 < \kappa \leq 1$, and any positive value of $\kappa$ has $\Atil \geq 0.055$, which is less than $1\sigma$ below the observed value if $\Delta\bar\sigma_*/\bar\sigma_* \simeq 1$.  Negative values of $\kappa$ are less promising; $\kappa=-1$ maximizes $K_\ell$ for negative $\kappa$, and it gives $\Atil=0.043$ for $\ell_\mathrm{max}=64$.  We will therefore only consider positive values for $\kappa$ for the rest of the analysis.  From Eq.~(\ref{tssCase2}) we see that $\kappa$ can be positive only if $\tilde{B} \propto B_\mathrm{m}$ is greater than $1/2$.  We are therefore only considering curvaton scenarios in which the curvaton at least partially decays into dark matter, but its contribution to the dark matter density is small compared to what existed prior to curvaton decay.  

\begin{figure}
\includegraphics[width=8.5cm]{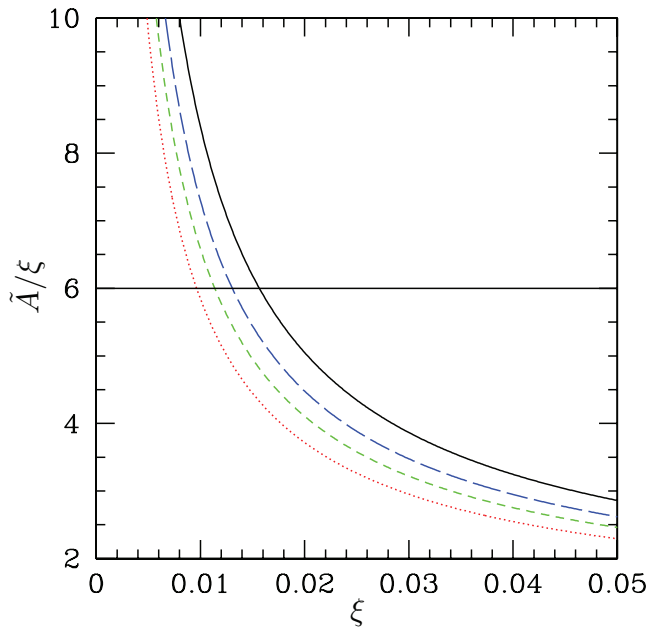}
\caption{The ratio $\Atil/\xi$, with $\ell_\mathrm{max}=64$, as a function of $\xi$ for four values of $\kappa^2\xi$: $\kappa^2\xi = 0.0086$ (solid), 0.007 (long-dashed), 0.006 (short-dashed), and 0.005 (dotted).  Since $K_\ell \rightarrow \xi$ as $\ell \rightarrow \infty$, this ratio illustrates the fractional enhancement in the asymmetry on large scales ($\ell \leq 64$) compared to small scales.   Since we require the asymmetry to be about 6 times larger on large scales than on small scales, we see that we require $\xi \lsim 0.016$.}
\label{fig:Klratio}
\end{figure}

Figure~\ref{fig:Klcase2} also illustrates how $\xi$ determines the small-scale value of $K_\ell.$  As $\xi$ decreases and $\kappa^2$ increases, $K_\ell$ decreases on small scales, and we see that $\kappa=-1$ and $\kappa=1$ give the same small-scale value for $K_\ell$.  Furthermore, the ratio of $K_\ell$ on large scales to $K_\ell$ on small scales decreases with increasing $\xi$.   We want the asymmetry to go from $A \simeq 0.072$ on large scales ($\ell_\mathrm{max} = 64$ in the CMB)  to $A \lsim 0.012$ on small scales ($k \simeq 1.3h-1.8h$ Mpc).  As mentioned above, the isocurvature perturbations' contribution to the total power is negligible on small scales, and any asymmetry is due solely to adiabatic perturbations from the curvaton, which implies that $A = \xi (\Delta\bar\sigma_*/\bar\sigma_*)$ on these scales.  A reduction in $A$ from 0.072 on large scales to less than 0.012 on small scales therefore requires that $\Atil/\xi \gsim 6$.  This requirement places an upper bound on $\xi$, as shown in Fig.~\ref{fig:Klratio}.  For $\kappa^2\xi = 0.0086$, which saturates the upper bound on isocurvature power, we see that the required enhancement on large scale is attained only if $\xi \lsim 0.016$, and the upper bound on $\xi$ decreases with decreasing $\alpha$.  This upper limit implies that the curvaton contributes only a small fraction of the adiabatic power (although it is a much bigger fraction than in the $\tss \simeq 1$ case).  The adiabatic and isocurvature fluctuations are therefore nearly uncorrelated.

\begin{figure}
\includegraphics[width=8.5cm]{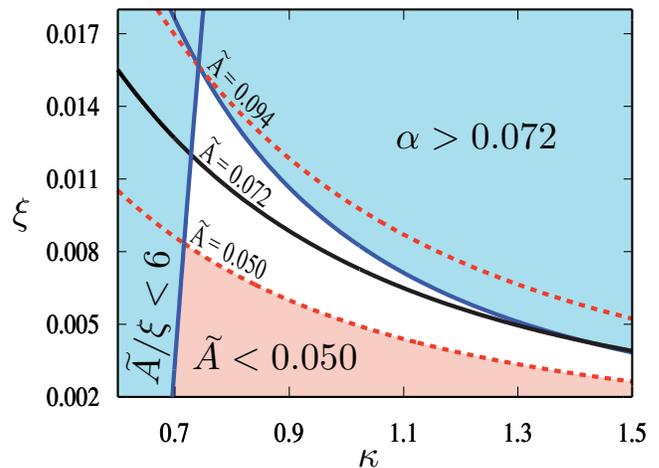}
\caption{The $\xi-\kappa$ parameter space for models in which the curvaton does not contribute significantly to the dark matter density.  The shaded region in the upper-right corner is excluded by the upper bound on isocurvature power ($\alpha<0.072$), and the left shaded region is excluded by the scale-invariance of the resulting asymmetry ($\Atil/\xi < 6$).  The dotted curves show where the maximal possible asymmetry $\Atil$ equals the observed asymmetry $\pm1\sigma$; the bottom shaded region cannot produce an asymmetry within $1\sigma$ of the observed value.}
\label{fig:contour}
\end{figure}

We now see that the required scale-dependence of the asymmetry limits its magnitude: for fixed $\alpha$, the asymmetry is maximized if $\xi$ is large and $\kappa$ is small, but increasing $\xi$ makes the asymmetry more scale-invariant.   Figure \ref{fig:contour} summarizes the constraints on $\kappa$ and $\xi$.  We see that only a limited region of the $\kappa$-$\xi$ plane can produce asymmetries with $A\geq 0.072$ while also satisfying the upper bound on isocurvature power ($\alpha<0.072$) and the quasar constraint ($\Atil/\xi > 6$).  Since the quasar constraint is effectively an upper bound on $\xi$ and a lower bound on $\kappa$, a tighter constraint on asymmetry in the distribution of quasars could rule out models with $A\gsim0.072$; the allowed $\Atil \gsim 0.072$ region is excluded if we require that $\Atil/\xi >16.4$, which is a factor of 2.7 improvement on the current quasar constraint.  If future observations reveal that $A\lsim 0.055$, however, then the allowed range of $\kappa$ values has no upper bound, and it will not be possible to rule out these models by tightening the quasar constraint.

In contrast, lowering the upper bound on the isocurvature fraction $\alpha$ will always decrease the asymmetry amplitudes that are accessible to our model. 
The minimum values of $\alpha$ in the $\Atil \geq 0.072$ and $\Atil \geq 0.050$ allowed regions are 0.054 and 0.037, respectively.  If there are no isocurvature modes, then the Planck satellite should give a 95\% upper limit on isocurvature power of $\alpha<0.042$, while an ideal, cosmic-variance-limited CMB temperature and polarization map up to $\ell = 2000$ would constrain $\alpha<0.017$ \cite{HHRW09}.  Thus we see that Planck is capable of ruling out our model if $A\gsim0.056$ for $\ell\lsim64$, and subsequent CMB measurements could test models with even smaller asymmetry ($\Atil \gsim 0.023$).  

The maximum value for $\Atil$ is obtained when both the upper bound on $\alpha$ and the upper bound on $\xi$ are saturated; as shown in Fig.~\ref{fig:contour}, $\Atil = 0.094$ for $\xi=0.016$ and $\kappa= 0.74$.  Since $A=(\Delta\bar\sigma_*/\bar\sigma_*)\Atil$, we see that $A \gsim 0.05$, which is $1\sigma$ below the observed value, can only be obtained if $(\Delta\bar\sigma_*/\bar\sigma_*) \gsim 1/2$.  With this lower bound on $(\Delta\bar\sigma_*/\bar\sigma_*)$, the CMB dipole and quadrupole constraints given by Eqs. (\ref{dipole}) and (\ref{quad}) become upper bounds on $R$ that are inversely proportional to $\kappa$.  With ${\cal D} =0.0034$ and $Q=1.9\times10^{-5}$, the CMB quadrupole provides a stronger constraint in the allowed region shown in Fig.~\ref{fig:contour} ($\kappa\gsim 0.7$); for $\kappa \simeq 1.4$, $R\lsim0.00013$ is required to satisfy the CMB constraints.  Since $K_\ell$ does not depend on $R$ if $\tss = \kappa R$, it will be possible to evade these constraints without changing the asymmetry.  (Even though $\xi$ depends on $R$, we can treat $\xi$ and $R$ as independent variables because $\xi$ also depends on $\eh$ and $R$ does not.)  

The upper limit on $R$ does have consequences for the non-Gaussianity parameters, however.  
From Eq. (\ref{fnlad}) for $f_\mathrm{NL}$ we see that the upper bound on $R$ and the constraint  $f_\mathrm{NL}\lsim 80$ implies that $\xi\lsim0.093$.  Since this upper bound is much larger than the $\xi$ values required to generate the necessary suppression of the power asymmetry on small scales, non-Gaussianity in the adiabatic perturbations is not a concern.  The non-Gaussianity from the isocurvature perturbations is bounded from below by the CMB dipole constraint; Eq.~(\ref{dipole}) implies an upper bound on $\tss$ that leads directly to a lower bound on $f_\mathrm{NL}^\mathrm{iso}$ through Eq.~(\ref{fnliso}).  With the maximal variation in $\bar\sigma_*$ ($\Delta\bar\sigma_*/\bar\sigma_* =1)$, the CMB dipole constraint implies that $f_\mathrm{NL}^\mathrm{iso} \gsim 82$, independent of $\kappa$.   If we restrict ourselves to $\kappa\lsim1.4$, then the upper bound on $R$ from the CMB quadrupole, $R\lsim0.00013$ for $\Delta\bar\sigma_*/\bar\sigma_*\gsim 1/2$, implies that $f_\mathrm{NL}^\mathrm{iso} \gsim 1300$.  If $\alpha = 0.072$, then these lower limits on $f_\mathrm{NL}^\mathrm{iso} $ correspond to $f_\mathrm{NL}\gsim 0.1$ and $f_\mathrm{NL}\gsim 2.3$, respectively \cite{HKMTY09}; both of these lower bounds on $f_\mathrm{NL}$ are well within current observational limits and are probably beyond the reach of the Planck satellite \cite{KS01,BZ04,Yadav08}.

Thus we see that this model for generating the power asymmetry does not require observable non-Gaussianity, unlike the purely adiabatic scenario described in Ref. \cite{EKC08}.  We also note that significant departures from Gaussianity, should be they be observed in the future, can be accommodated by this model by decreasing $R$.  This non-Gaussianity will include contributions from isocurvature perturbations, however, so its scale-dependence may be distinguishable from purely adiabatic non-Gaussianity \cite{KNSST08, HKMTY09, KNSST09}.

%%%%%%%%%%%%%%%%%
\section{Summary and Discussion}
\label{sec:conclusions}
%%%%%%%%%%%%%%%%%

A large-amplitude superhorizon fluctuation in the curvaton field can generate a hemispherical power asymmetry, provided that the curvaton is always a subdominant component of the Universe's energy density.  If the curvaton decays while all particle species are still in thermal equilibrium with radiation, then the fluctuations in the curvaton field create only adiabatic perturbations, and the resulting asymmetry is scale-invariant \cite{EKC08}.  Recent studies have revealed that the asymmetry is not scale-invariant; while $\Delta C_\ell /C_\ell \simeq 0.15$ for $\ell \lsim 64 $ \cite{Eriksen09}, there are indications that this asymmetry does not extend to $\ell \gsim 600$ \cite{Lew08, HBGEL08}, and an analysis of quasar number counts found $\Delta P(k)/P(k) \lsim 0.024$ for $k \simeq 1.3h-1.8h$ Mpc$^{-1}$.  With the aim of explaining this scale-dependence, we have considered how the asymmetry produced by a large-amplitude curvaton fluctuation changes if the curvaton decays after dark matter freezes out.  In this scenario, the curvaton produces dark-matter isocurvature perturbations in addition to adiabatic perturbations, and both types of perturbations have asymmetric power.  Since isocurvature fluctuations decay after entering the horizon, their contribution to the CMB power spectrum is much greater on large scales ($\ell \lsim 100$) than on smaller scales, and the magnitude of the asymmetry will decrease accordingly.

There are two limiting cases if the curvaton decays after dark matter freezes out: the majority of the dark matter can be created when the curvaton decays, or the curvaton's contribution to the dark matter density may be insignificant.  In the first scenario, the isocurvature fluctuations from the curvaton are much larger than the adiabatic perturbations from the curvaton, and all of the adiabatic power comes from inflaton fluctuations.   Since only the isocurvature fluctuations are asymmetric, it is very difficult to generate the observed asymmetry without violating the current bound on power from isocurvature modes.   It is necessary to introduce an order-unity variation in the curvaton density across the observable Universe, and since the curvaton creates the dark matter in this model, this would have profound observational consequences.  For instance, the resulting large-amplitude isocurvature perturbation induces a temperature dipole in the CMB that is far too large to be consistent with observations.  We conclude that it is not possible to generate the observed asymmetry with a superhorizon curvaton fluctuation if the curvaton creates the dark matter.

The second scenario, in which the curvaton's contribution to the dark matter density is negligible, is far more promising.  In this scenario, the curvaton produces adiabatic and isocurvature fluctuations of roughly equal amplitude.  It is therefore slightly easier to generate the observed asymmetry, but the requirement that the asymmetry magnitude decrease by a factor of 6 between large and small scales limits the curvaton's contribution to the total adiabatic power to less than 1.6\%.  Consequently, the variation in the curvaton field across the observable Universe must be greater than 50\% to generate the observed asymmetry.  Fortunately, the amplitude of both the isocurvature mode and the adiabatic mode generated by this superhorizon variation in the curvaton is proportional to the fraction $R$ of the total energy density contained in the curvaton at the moment of its decay.  We can therefore suppress any observational signature of the superhorizon curvaton fluctuation in the CMB without altering the asymmetry by decreasing the energy density of the curvaton.  Decreasing $R$ does increase the non-Gaussianity of the fluctuations created by the curvaton, but the resulting non-Gaussianity is well within the current observational bounds.   

We conclude that a superhorizon fluctuation in the curvaton field is capable of generating the observed asymmetry in the CMB while satisfying the upper bound on asymmetry in the quasar population if the curvaton's contribution to the dark matter is negligible.  The curvaton scenario employed by our model has two free parameters: $\xi$ is the fraction of the adiabatic power that comes from the curvaton field, and $\kappa$ determines the strength of the isocurvature perturbation created by the curvaton through $\smr = \kappa R \ssigr$.  Both $\xi$ and $\kappa$ depend on other features of the curvaton model; $\xi$ depends on the slow-roll parameter $\eh$, $R$, and the initial value of the curvaton field, while $\kappa$ depends on the fraction of curvaton energy that is converted to dark matter and the dark matter density at curvaton decay.  The asymmetry parameter $A = (1/2)\Delta C_\ell /C_\ell$ for the WMAP5 ILC CMB map is $0.072\pm0.022$ for $\ell\lsim64$.  We find that only a narrow region of the $\kappa$-$\xi$ parameter space is capable of generating an asymmetry with $A\gsim 0.072$, as shown in Fig.~\ref{fig:contour}, so our model requires a fair amount of fine-tuning.  If future observations reveal that $A\simeq0.05$, then the allowed region opens up considerably and includes $\xi\simeq0$ with $\kappa \gg 1$.  Negative values of $\kappa$ cannot generate the observed asymmetry, however, which implies that the curvaton must at least partially decay into dark matter. 

The observational consequences of our model for the origin of the power asymmetry differ considerably from the predictions of the purely adiabatic model proposed in Ref. \cite{EKC08}.  Whereas the purely adiabatic model requires $\xi \gsim 0.072$ to generate the observed asymmetry and can work with $\xi \simeq 1$, our model is constrained to much smaller values of $\xi$ ($\xi \lsim 0.016$).  Consequently, our model does not significantly change the tensor-scalar ratio [$r=16\eh(1-\xi)$] or the inflationary consistency relation.   The dominance of the inflaton's contribution to the primordial fluctuations also implies that our model does not require detectable levels of non-Gaussianity, unlike the purely adiabatic model which predicts $f_\mathrm{NL}\gsim 26$ for $A=0.072$.  There is one shared prediction, however; any primordial origin of the asymmetry predicts that the anisotropy in the power spectrum will produce signatures in the CMB polarization and temperature-polarization correlations \cite{DPH07}.  Our model also predicts a hemispherical asymmetry in the isocurvature fraction.

Finally, we note that this method of generating a scale-dependent power asymmetry through isocurvature perturbations produces an asymmetry with a specific spectrum.  The magnitude of the resulting asymmetry peaks at $\ell \simeq 10$, rapidly decreases for $\ell = 10-100$, and is nearly gone for $\ell\gsim100$.  Throughout this paper we have considered the scale-averaged asymmetry parameter $A$ with $\ell_\mathrm{max} = 64$; the observed value for this $\ell$ range in $V$-band is $A = 0.080\pm0.021$ \cite{Eriksen09}.  To probe the scale-dependence of the asymmetry, Ref. \cite{Eriksen09} also considered two other values of $\ell_\mathrm{max}$ in $V$-band and found that $A = 0.119\pm0.034$ for $\ell_\mathrm{max} = 40$ and $A = 0.070\pm 0.019$ for $\ell_\mathrm{max} =80$.   Despite the rapid fall-off of the asymmetry generated by our model for $\ell\gsim10$, it is consistent with these nearly scale-invariant results.  For instance, if $\kappa = 0.75$ and $\xi=0.013$ (a point near the middle of the allowed region in Fig.~\ref{fig:contour}),  then $A=0.113(\Delta\bar\sigma_*/\bar\sigma_*)$ for $\ell_\mathrm{max}=40$, $A=0.080(\Delta\bar\sigma_*/\bar\sigma_*)$ for $\ell_\mathrm{max}=64$, and $A=0.065(\Delta\bar\sigma_*/\bar\sigma_*)$ for $\ell_\mathrm{max}=80$.   Moreover, a search for asymmetry in the amplitude of the first acoustic peak found that $A<0.03$ at the 95\% C.L. for $\ell\simeq220$ \cite{DD05}, which is consistent with the predictions of our model. 

There are also indications that the asymmetry is present, at least to some extent, out to $\ell \simeq 600$ \cite{HBGEL08}.  Unfortunately, the asymmetry parameterization employed in this analysis cannot be directly related to the $A$ parameter, so it is difficult to interpret these results.  An analysis analogous to Refs. \cite{Eriksen07, Eriksen09} out to higher $\ell$ values is required to determine whether the scale-dependence of $A$ predicted by our model is consistent with observations.   Our model also predicts that at least $5.4\%\,(3.7\%)$ of the primordial power comes from isocurvature fluctuations if $A\gsim0.072\,(0.050)$ for $\ell\lsim64$.  Future searches for isocurvature fluctuations will therefore provide an additional test of our proposed origin of the CMB hemispherical power asymmetry; in particular, our model predicts that the Planck satellite should detect isocurvature perturbations if $A\gsim0.056$ for $\ell\lsim64$ \cite{HHRW09}.  

\begin{acknowledgments}
We thank Hans Kristian Eriksen for helpful correspondence.    AE and MK are
supported by DoE DE-FG03-92-ER40701 and the Gordon and
Betty Moore Foundation.  CH is supported by DoE DE-FG03-02-ER40701, the National Science Foundation under contract AST-0807337, and the Alfred P. Sloan Foundation.
\end{acknowledgments}

%%%%%%%%%%%%%%%%%%%%%%%%%%%%%
\appendix
\section{A scale-dependent asymmetry without isocurvature modes?}
\label{app:varyingXi}
%%%%%%%%%%%%%%%%%%%%%%%%%%%%%
In this appendix, we will attempt to make the power asymmetry generated by a superhorizon curvaton fluctuation scale-dependent without introducing isocurvature perturbations.  As in Ref. \cite{EKC08}, we define $\xi$ to be the fraction of the total power that comes from the curvaton ($\sigma$):
\beq
\xi(k) = \frac{{\cal P}_{\zeta,\sigma}(k)}{{\cal P}_{\zeta,\sigma}(k)+{\cal P}_{\zeta,\phi}(k)},
\label{xidef}
\eeq
where $\phi$ refers to the inflaton and $\zeta$ is the curvature of uniform-density hypersurfaces defined by Eq.~(\ref{zetadef}).  The adiabatic power asymmetry amplitude is $\Delta P(k)/P(k) = 2\xi(\Delta\bar\sigma/\bar\sigma)$.  Since $\bar\sigma$ has no scale dependence, the only way to make $\Delta P(k)/P(k)$ dependent on $k$ is to make $\xi$ dependent on $k$. We need $\Delta P(k)/P(k)$ to go from $2A = 0.144$ on large CMB scales ($\ell\lsim64$) \cite{Eriksen09} to less than 0.024 on quasar scales ($k \simeq 1.5h$ Mpc$^{-1}$) \cite{Hirata09}.  

We recall from Section \ref{sec:cmb} that
${\cal P}_{\zeta,\phi}(k) = \Asqrd$,
and
${\cal P}_{\zeta,\sigma}(k) = (R/3)^2 \Bsqrd$, where $\Asqrd$ and $\Bsqrd$ are defined by Eqs. (\ref{Asqrd}) and (\ref{Bsqrd}).
Inserting these expressions into Eq.~(\ref{xidef}) gives
\beq
\xi(k) = \left[ 1+ \frac{9\pi}{R^2\eh(k)}\left(\frac{\bar\sigma_*}{\mpl}\right)^2\right]^{-1},
\eeq
where $\mpl^2 = G^{-1}$.  Thus we see that any scale-dependence in $\xi$ must originate from variation in the slow-roll parameter $\eh$ during inflation, and we note that $\eh$ depends on the inflaton potential through \cite{LPB94}
\beq
\eh(k) \simeq \ev(k) \equiv \frac{\mpl^2}{16\pi}\left.\left[\frac{V^\prime(\phi)}{V(\phi)}\right]^2\right|_{k=aH(\phi)}.
\eeq
It will be useful to define
\beq
\et(k) \equiv \frac{1}{9\pi} \left(\frac{\bar\sigma_*}{\mpl}\right)^{-2} R^2 \eh(k)
\eeq
because then $\xi$ has a simple form:
\beq
\xi(k) = \frac{\et}{\et+1}.
\label{xiwithet}
\eeq
Even though $R\ll1$ and $\eh \ll 1$, $\et$ may take any value; $\et \gg 1$ is possible if $\bar\sigma \ll \mpl$, while $\et \ll 1$ can be obtained by decreasing $R$ or $\eh$.  

There are two ways that we can give $\xi$ the scale-dependence necessary to generate the observed asymmetry in the CMB and still satisfy the quasar constraint on small-scale asymmetry: we can make $\xi$ discontinuous by inserting a kink into $V^\prime(\phi)/V(\phi)$, or we can choose the spectral indices of ${\cal P}_{\zeta,\sigma}(k)$ and ${\cal P}_{\zeta,\phi}(k)$ in such a way that $\xi(k)$ decreases smoothly as $k$ increases.  We will consider both approaches in this appendix.

%%%%%%%%%%%%%%%%%%%%%%%
\subsection{A discontinuity in $\xi(k)?$}
\label{sec:step}
%%%%%%%%%%%%%%%%%%%%%%%

First, we will examine inflation models with broken scale-invariance.   We suppose that $\xi$ is a step function: $\xi = \ximax$ for $k<k_*$ and $\xi = \ximin$ for $k>k_*$ where both $\ximax$ and $\ximin$ are constants.  To achieve the necessary suppression of the asymmetry on small scales, we require that $\ximin/\ximax \lsim 1/6$ and $k_*$ must be located somewhere between large CMB scales ($k\simeq0.0033h$ Mpc$^{-1}$) and quasar scales ($k \simeq 1.5h$ Mpc$^{-1}$) \cite{Hirata09}.  
A step function in $\xi$ requires a step function in $\et$, which requires a step function in $\eh$.  From $\ximin/\ximax = 1/6$ and Eq.~(\ref{xiwithet}), we see that 
\beq 
\frac{\etmin}{\etmax} = \frac{\ehmin}{\ehmax} = \frac{1}{6+5\etmax}.
\label{xeqn}
\eeq
Thus we see that the size of the necessary discontinuity in $\eh$ is determined by $\ximax$.  If the curvaton dominates on large scales ($\xi \simeq 1$ and $\et \gg 1$), then the drop in $\eh$ necessary to make the inflaton dominant on small scales is large.  This drop is minimized if $\ximax$ is minimized, but $\ximax \gsim 0.07$ is required to generate the observed asymmetry \cite{EKC08}.

Now we have to worry about the shape of the power spectrum.  Inserting a downward step in $\eh$ leaves the curvaton perturbation spectrum unaltered, but it gives the inflaton perturbation spectrum an upward step.  Therefore, we would expect that the total power on large scales would be smaller than the total power on small scales.   In contrast, the primordial ${\cal P}_\zeta(k)$ that fits CMB and large-scale-structure observations is nearly flat on all scales from $k=0.0001$ Mpc$^{-1}$ to $k=0.2$ Mpc$^{-1}$ \cite{VP08}.  
Furthermore, the value for $\sigma_8$, the fluctuation amplitude at 8$h$ Mpc$^{-1}$, derived from the CMB is consistent with the measurements from weak lensing observations \cite{WMAP5only}, so there can no major change in the primordial power spectrum between the scales probed by the CMB and those probed by weak lensing.  We conclude that the total primordial power spectrum must be nearly scale-invariant.

The total primordial power spectrum may be expressed in terms of $\xi$ and the curvaton power:
\beq 
{\cal P}_\zeta(k) = \frac{1}{\xi(k)}\left(\frac{R}{3}\right)^2\frac{H_\mathrm{inf}^2}{\pi^2\bar\sigma_*^2}.
\eeq
The ratio of power on large scales $(k_\mathrm{CMB}<k_*)$ to the power on small scales $(k_\mathrm{Q}>k_*)$ is 
\beq
\frac{{\cal P}_\zeta(k_\mathrm{CMB})}{{\cal P}_\zeta(k_\mathrm{Q})} = \left[\frac{H_\mathrm{inf}^2(k_\mathrm{CMB})}{H_\mathrm{inf}^2(k_\mathrm{Q})}\right]\left(\frac{\ximin}{\ximax}\right).
\label{powerratio}
\eeq
To compensate for the injection of inflaton power on scales smaller than $k_*$, we must introduce a discontinuity in $V(\phi)$ at the same $\phi$ value as the discontinuity in $V'(\phi)$.  Since $H^2_\mathrm{inf} \propto V(\phi)$, we see that 
\beq
\frac{V(\phi_\mathrm{CMB})}{V(\phi_\mathrm{Q})} = \frac{\ximax}{\ximin} \gsim 6,
\eeq
where $k_\mathrm{CMB,Q} = a H_\mathrm{inf}(\phi_\mathrm{CMB,Q})$, is required to keep the primordial power spectrum scale-invariant.  

Thus we see that it is possible to hide the kick that the power spectrum gets when the inflaton takes over on small scales by introducing a drop in the inflationary energy scale that leads to a total reduction in power in both the curvaton and inflaton fluctuations.  As the inflaton rolls across the discontinuity in the power spectrum, the value of $V(\phi)$ must drop by at least a factor of 6, and the potential on the lower side must be significantly flatter than the potential on the upper side; from $\eh \propto (V^\prime(\phi)/V)^2$ and Eq.~(\ref{xeqn}), we conclude that
\beq
\frac{V^\prime(\phi_\mathrm{Q})}{V^\prime(\phi_\mathrm{CMB})} =  \frac{1}{6}\sqrt{\frac{1}{6+5\etmax}}
\eeq
for $\ximin/\ximax = 1/6$.  
Note that there is no lower bound on $\eh$ on either side of the potential break; $V(\phi)$ can be as flat as we want it to be on large scales provided that it is even flatter on small scales.  Therefore, we do not have to worry about constraints from the tensor-scalar ratio or the scalar spectral index.  

There is another concern, however.  Even though the discontinuities in $V(\phi)$ and $V^\prime(\phi)$ conspire to preserve the flatness of the total power spectrum, the momentary interruption of slow-roll inflation that occurs as the inflaton field crosses the break could induce oscillations in the power spectrum localized around $k_*$.  Ref. \cite{Covi06} analyzes the effects of a step in the mass $m$ of a quadratic $V=m^2\phi^2/2$ potential, and Ref. \cite{HCMS07} generalizes this analysis to steps in other inflaton potentials.  They restrict their analyses to breaks in the potential with $\Delta V/\bar{V} \leq 0.2$ because they do not want the inflaton's kinetic energy to exceed its potential energy after the inflaton crosses the break.  Even with this constraint, the smoothness of the observed power spectrum is very restrictive; CMB and large-scale-structure observations constrain $\Delta V/\bar{V} \lsim 10^{-3}$ at 99\% confidence for $k_* \lsim 0.1$ Mpc$^{-1}$ \cite{HCMS07}.

The broken-scale-invariance models considered in Refs. \cite{Covi06, HCMS07} include a small change in the perturbation amplitude across the step and do not include perturbations from a curvaton field, but their findings are still very discouraging for our proposal.  They find that the amplitude of the oscillations far exceeds the change in ${\cal P}_\zeta(k)$ across the break, so it is reasonable to expect that any break in the potential will induce large oscillations, even if ${\cal P}_\zeta(k)$ is unaffected.  Furthermore, the oscillations reach their maximum on scales that leave the horizon after $\phi$ crosses the break.  On these scales, the inflaton must dominate the power spectrum to suppress the asymmetry, so there is no hope of masking the oscillations in the inflaton spectrum with the curvaton spectrum.  We could hope to hide the oscillations induced by our model by setting $k_* \gsim 0.1$ Mpc$^{-1}$, but the potential change required by our model is so large that inflation may not resume after the inflaton crosses the break.  We therefore conclude that the discontinuity in $V(\phi)$ that is required by the broken scale-invariance model to satisfy the quasar constraint is highly unlikely to be consistent with observations.

%%%%%%%%%%%%%%%
\subsection{A smooth transition?}
\label{sec:smooth}
%%%%%%%%%%%%%%%%%%

Next we consider the possibility that $\xi$, and therefore $\eh$, smoothly decrease as $k$ increases.  Given the difference in $k$ between large CMB scales and quasar scales, we would need
\beq
\frac{\drm \ln \xi}{\drm \ln k} \simeq \frac{\Delta \ln \xi}{\Delta \ln k} \simeq - \frac{1.8}{6.1}  = -0.29
\eeq
if (${\drm \ln \xi}/{\drm \ln k}$) is to be roughly constant over the scales of interest ($0.0033h \,\mathrm{Mpc}^{-1} \lsim k \lsim1.5h \, \mathrm{Mpc}^{-1}$).

The spectral index for $\xi$ is related to the spectral indices of the power spectrum for inflaton and curvaton fluctuations \cite{LW02,ISTY08}:
\beqa
\frac{\drm \ln \xi}{\drm \ln k} &=& \frac{\drm \ln {\cal P}_{\zeta,\sigma}}{\drm \ln k} - \xi \frac{\drm \ln {\cal P}_{\zeta,\sigma}}{\drm \ln k} - (1-\xi) \frac{\drm \ln {\cal P}_{\zeta,\phi}}{\drm \ln k} \nonumber\\
&=& -2\eh - \xi(-2\eh)-(1-\xi)(-4\eh+2\nh)\nonumber\\
&=& -(1-\xi)(2\nh-2\eh)
\eeqa
 where $\nh$ is the slow-roll parameter
\beq
\nh \equiv -\frac{\ddot{\phi}}{\dot{\phi}H},
\eeq
and we note that $\nh \simeq (\mpl^2/8\pi)[V^{\prime\prime}(\phi)/V(\phi)] - \eh$ \cite{LPB94}.  Meanwhile, the spectral index for the total power spectrum is
\beqa
n_s -1 &\equiv& \frac{\drm \ln [{\cal P}_{\zeta,\sigma}+{\cal P}_{\zeta,\phi}]}{\drm \ln k}, \\
&=& -2\eh-(1-\xi)(2\eh-2\nh) \label{ns},
\eeqa
and the tensor-scalar ratio is
\beq
r=16\eh(1-\xi).
\eeq
Therefore, the spectral index for $\xi$ depends only on $\xi$ and observable parameters:
\beq 
-\frac{\drm \ln \xi}{\drm \ln k} = n_s-1+\frac{r}{8(1-\xi)} 
\label{bound1}.
\eeq

The scalar spectral index $n_s$ and the tensor-scalar ratio ratio $r$ at $k = 0.002$ Mpc$^{-1}$ are degenerate parameters; the blue tilt introduced by making $n_s>1$ may be compensated for by increasing $r$, which adds power on large scales.  When $n_s$ is assumed to be scale-invariant, the 2$\sigma$ upper limits on $n_s$ and $r$ from WMAP5+BAO+SN are $n_s \lsim 1.01$ and $r \lsim 0.22$ \cite{WMAP5}.  The introduction of running in the scalar spectral index brings even larger values of $n_s$ and $r$ into the $2\sigma$-allowed region of parameter space, but the allowed range for the running index,
$\alpha_s \equiv \drm n_s/\drm \ln k$,
is negatively correlated with $r$ and $n_s$.  This correlation makes it difficult to obtain a large negative value for ${\drm \ln \xi}/{\drm \ln k}$ because $\alpha_s$ is also connected to the spectral index for $\xi$ through Eq.~(\ref{ns}). 

To evaluate $\alpha_s$, we will need the running of the slow-roll parameters
\beqa
 \frac{\drm \ln \eh}{\drm \ln k} &=& 2(\eh-\nh), \\
 \frac{\drm \ln \nh}{\drm \ln k} &=& \eh - \frac{\xi_H^2}{\nh},
 \eeqa
 where 
 \beq
 \xi_H \equiv \frac{\mpl^2}{4\pi}\sqrt{\frac{H^\prime(\phi)H^{\prime\prime\prime}(\phi)}{H^2}}
 \eeq
 is a higher-order slow-roll parameter \cite{LPB94}.  From Eq.~(\ref{ns}), it follows that
 \beq
 -\frac{\drm \ln \xi}{\drm \ln k} = \frac{\alpha_s - (r/8)\left(\nh -  {\xi_H^2/\eh}\right)}{\frac{\xi(n_s-1)}{1-\xi}+\frac{r/4}{(1-\xi)^2}}.
\label{bound2}
\eeq  
Since $\xi \leq 1$ by definition, demanding that the right-hand-side of Eq.~(\ref{bound1}) be positive implies that the denominator on the right-hand-side of Eq.~(\ref{bound2}) is positive.  The combination of slow-roll parameters, $\nh-\xi_H^2/\eh$, in the numerator could be positive or negative, but $|\nh-\xi_H^2/\eh| \ll 1$ is required by the slow-roll approximation.  If $\alpha_s$ is negative then the right-hand-side of Eq.~(\ref{bound2}) can be positive only if $|\alpha_s|\ll r/8$, which forces $\alpha_s \simeq 0$.  We conclude that $\alpha \gsim 0$ is required to make ${\drm \ln \xi}/{\drm \ln k}$ negative, and this constraint makes the largest allowed values for $n_s$ and $r$ inaccessible.

To probe the possible evolution of $\xi$ during inflation, we derive a differential equation for $\xi(k)$.  We start with
\beq
-\frac{\drm \ln \xi}{\drm \ln k} = n_s(k)-1 + 2\eh(k).
\eeq
This equation is simply Eq.~(\ref{bound1}), but with $\eh$ instead of $r$.  From Eq.~(\ref{xiwithet}) it follows that
\beq
\eh(k) = \en \frac{\xi(k)}{1-\xi(k)},
\label{eh}
\eeq
where $\en \equiv \eh(k)/\et(k)$; $\en$ is therefore independent of $k$ and can take any value.   Since there are measurements of $n_s$ and its running, we will use these values for $n_s(k)$.  The resulting differential equation for $\xi(k)$ is
\beq
-\frac{\drm \ln \xi}{\drm \ln k} = n_s(k_0)-1+\alpha_s\ln \left(\frac{k}{k_0}\right) + 2\en \frac{\xi(k)}{1-\xi(k)}.
\label{Xidiffeq}
\eeq
As mentioned above, the 2$\sigma$ upper bound on $\alpha_s$ from WMAP5+BAO+SN is negatively correlated with $n_s$.  We therefore consider several points, subject to the constraint $\alpha_s>0$, on the boundary of the 2$\sigma$ error ellipse in $(n_s,\alpha_s)$ space (see Fig. 4 in Ref. \cite{WMAP5}) when choosing values for $n_s(k_0 = 0.002 \,\mbox{Mpc}^{-1})$ and $\alpha_s$.  Our aim is to minimize  the ratio  $\ximin/\ximax = \xi(1.5h\, \mathrm{Mpc}^{-1})/\xi(0.0033h\, \mathrm{Mpc}^{-1})$, which must be less than $1/6$ to satisfy the constraint on small-scale asymmetry from quasars.  We find that  $\ximin/\ximax$ is minimized if we choose the maximal allowed value for $n_s$, $n_s(k_0 = 0.002 \,\mbox{Mpc}^{-1}) = 1.03$, which corresponds to $\alpha_s=0$.  With this parameter choice, the right-hand-side of Eq.~(\ref{Xidiffeq}) is positive, so $\xi$ decreases monotonically as $k$ increases.

The only free parameters that remain are $\en$ and the ``initial" value of $\xi_s \equiv \xi(k_s)$, where $k_s$ is the smallest $k$ value of interest and the starting point of the numerical integration of Eq.~(\ref{Xidiffeq}).  Together, $\en$ and $\xi_s$ determine $\eh(k_s)$.  From Eq.~(\ref{eh}), we see that $\eh$ decreases as $\xi(k)$ decreases; since $\xi(k)$ is monotonically decreasing for all $k$, $\eh(k_s)$ is the maximum value that $\eh$ will attain in the scale range of interest.  Since we wish to maximize $-{\drm \ln \xi}/{\drm \ln k}$, we want to choose the largest possible value for $\en$.  There are two upper bounds to consider.  First, $\eh \ll 1$ is required by the slow-roll approximation.  If we insist that $\eh \leq 0.5$, then the maximum possible value for $\en$ is 
\beq
\en = 0.5 \frac{1-\xi_s}{\xi_s}.
\label{srmaxen}
\eeq
There is, however, an additional constraint.  Since $\eh(k)$ is a monotonically decreasing function of $k$, the value of $\eh(k_s>k_0)$ sets a lower bound on the value of $\eh(k_0)$, provided that $\eh$ is continuous.  There is an upper bound on $\eh(k_0)$ that follows from the measured upper bound on the tensor-to-scalar ratio $r$ evaluated at $k_0$: $r(k_0)<0.22$ at 95\% C.L. for $\alpha_s = 0$ \cite{WMAP5}. This upper bound implies that the maximum value of $\en$ should be
\beq
\en = \frac{0.014}{\xi_s},
\label{rmaxen}
\eeq
which is lower than Eq.~(\ref{srmaxen}) for $\xi_s \lsim 0.97$.    

We integrate Eq.~(\ref{Xidiffeq}) to obtain $\xi(k)$ for different values of $\xi_s$ with $\en$ given by both Eqs.~(\ref{srmaxen}) and (\ref{rmaxen}).  We set $k_s = 0.002$ Mpc$^{-1}$, and for each value of $\en$, we find the value of $\xi_s$ that gives the smallest value of $\ximin/\ximax = \xi(1.5h\, \mathrm{Mpc}^{-1})/\xi(0.0033h\, \mathrm{Mpc}^{-1})$ with $h=0.7$.  We find that it is possible to obtain 
$\ximin/\ximax \lsim 1/6$ only if we violate the condition that $r(k_0)<0.22$. 
In that case, $\ximin/\ximax \lsim 1/6$ if $\xi_s \lsim 0.25$ and $\en$ is given by Eq.~(\ref{srmaxen}).  It is encouraging that this range includes values for $\xi_s$ that are greater that 0.07 because this means that $\xi$ on large scales can be large enough to generate the observed asymmetry.  The downside is that this model has $\eh \simeq 0.5$ on large scales; given that $\xi_s \leq 0.25$, this means that $r\gsim6$ on the largest scales for which this model applies.  Since $\eh$ is a monotonically decreasing function of $k$, it is not possible to make $\eh$ small enough to satisfy $r < 0.22$ on the scales for which that bound applies $(k\simeq k_0)$ and then have it smoothly increase to the value necessary to give sufficient variation in $\xi$.  If we use Eq.~(\ref{rmaxen}) to force $r(k_s)<0.22$, then the minimal value for $\ximin/\ximax$ is only 0.56; this factor of 2 reduction in the asymmetry between large and small scales is insufficient to satisfy the quasar constraint.

In summary, it is possible for $\xi$ to be greater than 0.07 on large CMB scales and then smoothly decrease by a factor of six between large CMB scales and quasar scales in a way that keeps the total power spectrum flat enough to be consistent with observations.  However, the required values of $\xi$ and $\eh$ on the largest scales are inconsistent with the upper bound on the tensor-to-scalar ratio on these scales.  In order to satisfy this bound, we would have to discontinuously change the values of $\xi$ and $\eh$ to suppress the tensor-scalar ratio on large scales, and we saw in the previous section that such discontinuities are problematic. 

%%%%%%%%%%%%%%%%%%%%%%%%%%%%%
\section{The Derivation of $\trs$ and $\tss$ in the curvaton scenario}
\label{app:isocurv}
%%%%%%%%%%%%%%%%%%%%%%%%%%%%%
In this appendix, we briefly review the derivations of $\trs$ and $\tss$ in the curvaton scenario.  In the limit of instantaneous curvaton decay, the total curvature perturbation cannot change during the decay of the curvaton.  We can therefore obtain $\trs$ by equating $\zeta^{(\mathrm{f})}$, which is evaluated just after curvaton decay, to $\zeta^{(\mathrm{bd})}$, which is evaluated just prior to curvaton decay.  From Eq.~(\ref{zetadef}) we have
\beq 
\zeta^{(\mathrm{f})} = \zeta^{(\mathrm{i})} + \left[\frac{R}{3} + \left(\frac{\Omega_\mathrm{cdm}}{4\Omega_\gamma+3\Omega_\sigma+3\Omega_\mathrm{cdm}}\right)^{(\mathrm{bd})}{\cal T}_\mathrm{fr}\right]\ssigr.
\label{zetafinal}
\eeq
We will see below that $\zeta_\mathrm{cdm}^{(\mathrm{bd})}$ is not equal to its initial value $\zeta_\mathrm{cdm}^\mathrm{(i)} = \zeta^{(\mathrm{i})}$, and we have defined ${\cal T}_\mathrm{fr}$ through
\beq 
\zeta_\mathrm{cdm}^\mathrm{(bd)} = \zeta_\mathrm{cdm}^\mathrm{(i)} + \frac{{\cal T}_\mathrm{fr}}{3} \ssigr.
\label{freeze-out}
\eeq
Since the Universe must be radiation-dominated after curvaton decay, the assumption that $R\ll1$ also implies that $\Omega_\mathrm{cdm}^\mathrm{(bd)}\ll1$.  Furthermore, we will see that ${\cal T}_\mathrm{fr}\ll R$ if $R\ll1$.  We thus obtain Eq.~(\ref{trs}) for $\trs$.

If the dark matter freezes out prior to curvaton decay, then the dark-matter isocurvature perturbation is created in two stages.  The first stage occurs at dark matter freeze-out \cite{LW03}.  The abundance of dark matter after freeze-out is determined by the expansion rate at freeze-out, so immediately after freeze-out, the hypersurface of constant dark-matter density coincides with the hypersurface of constant total density.  In the presence of the curvaton field, this hypersurface is not a hypersurface of constant radiation density and a dark-matter isocurvature perturbation is created.  The change in $\zeta_\mathrm{cdm}$ during freeze-out is given by $\Delta\zeta_\mathrm{cdm} = ({{\cal T}_\mathrm{fr}}/{3}) \ssigr$ where
\beq 
{\cal T}_\mathrm{fr} = \frac{(\alpha - 3)\Omega_\sigma^{(\mathrm{fr})}}{2(\alpha - 2) + \Omega_\sigma^{(\mathrm{fr})}}.
\eeq
Since we assume that the Universe does not cease to be radiation dominated prior to curvaton decay,
\beq
\Omega_\sigma^{(\mathrm{fr})} =\Omega_\sigma^{(\mathrm{bd})}  \sqrt{\frac{H^{(\mathrm{bd})}}{H^{(\mathrm{fr})}}} \simeq \frac{4}{3} R\sqrt{\frac{\Gamma_\sigma}{1.4\,\Gamma_\mathrm{cdm}^{(\mathrm{fr})}}}.
\eeq
Between freeze-out and curvaton decay, the dark matter is non-interacting and $\zeta_\mathrm{cdm}$ is conserved; Eq.~(\ref{freeze-out}) therefore relates $\zeta_\mathrm{cdm}$ just prior to curvaton decay to its initial value.  In the case that $R\ll1$, ${\cal T}_\mathrm{fr} \ll (2/3) R$ since $\Gamma_\sigma \ll  \Gamma_\mathrm{cdm}^{(\mathrm{fr})}$ (dark matter freezes out prior to curvaton decay).

The second stage of dark-matter isocurvature creation occurs at curvaton decay.  The change in $\zeta_\mathrm{cdm}$ during curvaton decay is given by \cite{LM07}
\beq 
\zeta_\mathrm{cdm}^{(\mathrm{f})}-\zeta_\mathrm{cdm}^\mathrm{(bd)} = \frac{B_\mathrm{m}\Omega_\sigma^\mathrm{(bd)}}{\Omega_\mathrm{cdm}^\mathrm{(bd)}+B_\mathrm{m}\Omega_\sigma^\mathrm{(bd)}}\left[\zeta_\sigma^\mathrm{(bd)}-\zeta_\mathrm{cdm}^\mathrm{(bd)}\right].
\label{decay}
\eeq
We can relate $\smr$ to $\ssigr$ by combining Eq. (\ref{zetafinal}) for $\zeta^{(\mathrm{f})}\simeq\zr^{(\mathrm{f})}$ and Eqs.~(\ref{freeze-out}) and (\ref{decay}) for $\zeta_\mathrm{cdm}^{(\mathrm{f})}$; the result is Eq.~(\ref{tss}).  This derivation of $\tss$ neglected the possibility that the injection of dark matter particles at curvaton decay could raise the dark matter particle number density to the point that dark matter particles begin to self-annihilate again.  If dark matter annihilations resume after curvaton decay, then the final value of $\smr$ is suppressed.  It is possible to approximate the effect of a second stage of dark matter self-annihilation by multiplying $\tss$ by a factor of $1/(1+\Upsilon)$, where $\Upsilon \equiv \Gamma_\mathrm{cdm}/H$ evaluated just after curvaton decay \cite{LM07}.  Since we are most interested in curvaton models in which the curvaton produces very little dark matter, we will assume that $\Upsilon \ll1$.  We note, however, that our results can be easily adapted to cases where the self-annihilation of the dark matter after curvaton decay is significant.

\end{document}